\documentclass[preprint,12pt]{aastex}

\usepackage{graphicx}
\shortauthors{J{\o}rgensen et al.}

\begin{document}

\title{The Spitzer c2d Survey of Large, Nearby, Interstellar Clouds. \\ III. Perseus Observed with IRAC}

\author{Jes K. J{\o}rgensen\altaffilmark{1}, Paul M.
  Harvey\altaffilmark{2}, Neal J. Evans II\altaffilmark{2}, Tracy
  L. Huard\altaffilmark{1}, Lori E. Allen\altaffilmark{1}, Alicia
  Porras\altaffilmark{1}, Geoffrey A.  Blake\altaffilmark{3}, Tyler
  L. Bourke\altaffilmark{1}, Nicholas Chapman\altaffilmark{4}, Lucas
  Cieza\altaffilmark{2}, David W. Koerner\altaffilmark{5}, Shih-Ping
  Lai\altaffilmark{4}, Lee G. Mundy\altaffilmark{4}, Philip
  C. Myers\altaffilmark{1}, Deborah L. Padgett\altaffilmark{6}, Luisa
  Rebull\altaffilmark{6}, Anneila I.  Sargent\altaffilmark{7}, William
  Spiesman\altaffilmark{2}, Karl R.  Stapelfeldt\altaffilmark{8},
  Ewine F. van Dishoeck\altaffilmark{9}, Zahed Wahhaj\altaffilmark{5}
  \& Kaisa E. Young\altaffilmark{2}}

\altaffiltext{1}{Harvard-Smithsonian Center for Astrophysics, 60
  Garden Street MS42, MA 02138, jjorgensen@cfa.harvard.edu}

\altaffiltext{2}{Department of Astronomy, University of Texas at Austin,
  1 University Station C1400, Austin, TX~78712}

\altaffiltext{3}{Division of Geological and Planetary Sciences,
  MS~150-21, California Institute of Technology, Pasadena, CA~91125}

\altaffiltext{4}{Department of Astronomy, University of Maryland, College
  Park, MD~20742}

\altaffiltext{5}{Department of Physics and Astronomy, Northern Arizona
  University, NAU~Box~6010, Flagstaff, AZ~86011-6010}

\altaffiltext{6}{Spitzer Science Center, MC~220-6, California
  Institute of Technology, Pasadena, CA~91125}

\altaffiltext{7}{Division of Physics, Mathematics, and Astronomy,
  MS~105-24, California Institute of Technology, Pasadena, CA~91125}

\altaffiltext{8}{Jet Propulsion Laboratory, MS~183-900, California
  Institute of Technology, Pasadena, CA~91125}

\altaffiltext{9}{Leiden Observatory, PO Box 9513, NL~2300 RA Leiden,
  The Netherlands}

\begin{abstract}
We present observations of 3.86~degree$^2$ of the Perseus molecular
cloud complex at 3.6, 4.5, 5.8 and 8.0~$\mu$m with the Spitzer Space
Telescope Infrared Array Camera (IRAC). The maps show strong extended
emission arising from shocked H$_2$ in outflows in the region and from
polycyclic aromatic hydrocarbon (PAH) features, in particular in the
prominent Perseus ring. More than 120,000 sources are extracted toward
the cloud. Based on their IRAC colors and comparison to off-cloud and
extragalactic fields, we identify 400 candidate young stellar
objects. About two thirds of these are associated with the young
clusters IC~348 and NGC~1333, which account for 14\% of the surveyed
cloud area, while the remaining third is distributed over the
remaining cloud, including a number of smaller groups around B1, L1448
and L1455. We classify the young stellar objects according to the
traditional scheme based on the slope of their spectral energy
distributions from near-infrared 2MASS $K_s$ through mid-infrared IRAC
and MIPS 24~$\mu$m wavelengths. Significant differences are found for
the numbers of embedded Class I objects relative to more evolved Class
II objects in IC~348, NGC~1333 and the remaining cloud with the
embedded Class I and ``flat spectrum'' YSOs constituting 14\%, 36\%
and 47\% of the total number of YSOs identified in each of these
regions. These numbers suggest a difference in evolution over the
cloud: NGC~1333 previously has been suggested to be younger than
IC~348 based on near-infrared studies of the YSO populations and the
larger number of Class I objects in NGC~1333 supports this. The high
number of Class I objects in the extended cloud (61\% of the Class I
objects in the entire cloud) suggests that a significant fraction of
the current star formation is occuring outside these two
clusters. Finally we discuss a number of outflows and identify their
driving sources, including the known deeply embedded Class 0 sources
outside the two major clusters. The Class 0 objects are found to be
detected by Spitzer and have very red $[3.6]-[4.5]$ colors but do not
show similarly red $[5.8]-[8.0]$ colors. The Class 0 objects are
easily identifiable in color-color diagrams plotting these two colors
but are, in some cases, problematic to extract with automatic source
extraction routines due to the extended emission from shocked gas or
scattered light in cavities related to the associated outflows.
\end{abstract}


\section{Introduction}
Low-mass stars such as our own Sun are formed in a wide range of
different environments ranging from isolated cores to large, dense
clusters. To fully understand the differences in the formation
processes amongst these environments, large unbiased surveys and
compilations of young stellar objects are necessary. A major goal of
the Spitzer Space Telescope legacy project ``From Molecular Cores to
Planet Forming Disks'' \citep[c2d;][]{evans03} is to characterize star
formation in a variety of environments through a combination of
mapping and spectroscopic observations utilizing the full palette of
Spitzer instruments.

The Perseus molecular cloud is a prime example of a low-mass star
forming region. Low-mass protostars are found in a number of dense
clusters, such as NGC~1333 and IC~348, looser groups associated with
smaller dark clouds, including Barnard 1, Barnard 5, L1448 and L1455,
and a number of dense cores. Although no massive stars are currently
formed in the cloud, OB stars have been formed previously in the
vicinity, as witnessed by the Per OB2 association
\citep[e.g.,][]{dezeeuw99}. This paper presents Spitzer Infrared Array
Camera (IRAC) maps of 3.86~degree$^2$ of Perseus from the c2d legacy
project. This paper is the third in a series of papers presenting the
basic results for the five large clouds in the c2d survey. The two
previous papers in this series presented MIPS observations of
Chamaeleon \citep{young05} and IRAC observations of Serpens
\citep{harvey06serpens}. We will present the results for all clouds in
a fairly standard format to facilitate comparisons. While this paper
is based on IRAC data, some information from MIPS observations of
Perseus by c2d \citep{rebull06} is used in source classification and
in discussion of particular sources.

Perseus was first mapped, in full, in CO by
\cite{sargent79}. \cite{ladd93iras} identified 23 candidate young
stellar objects (YSOs) from the IRAS point source catalog and surveyed
these at 2~$\mu$m. \citeauthor{ladd93iras} found that the YSO
candidates were distributed throughout the cloud with higher numbers
toward the prominent clusters in IC~348 and NGC~1333: 87\% of the YSOs
were located within 1~degree of these two clusters, but only 30\%
within 30\arcmin. Subsequent near-infrared studies have focused on the
regions around NGC~1333 \citep{lada96,wilking04} and IC~348
\citep{lada95,luhman03}, identifying a large number of infrared excess
sources (about 60 sources in each region -- corresponding to 50\% and
20\% of the total number of cluster members in NGC~1333 and IC~348,
respectively).

The exact distance to Perseus remains uncertain. It is not clear
whether the cloud is located in the vicinity of the Per OB2
association (at a distance of 320~pc inferred from Hipparcos parallax
measurements \citep{dezeeuw99}) or whether the cloud is closer, in
front of the OB association at 200--250~pc such as suggested by
extinction studies \citep[e.g.,][]{cernis90}. Whether a single
distance to the entire cloud is valid or whether the Perseus cloud in
fact has a distance gradient (similar to an observed line of sight
velocity gradient \citep{sargent79}) or is a conglomeration of smaller
clouds is another issue. In any case, following \cite{enoch05}, a
distance of $250\pm 50$~pc is adopted in this paper.

\section{Observations}
\begin{figure}
\plotone{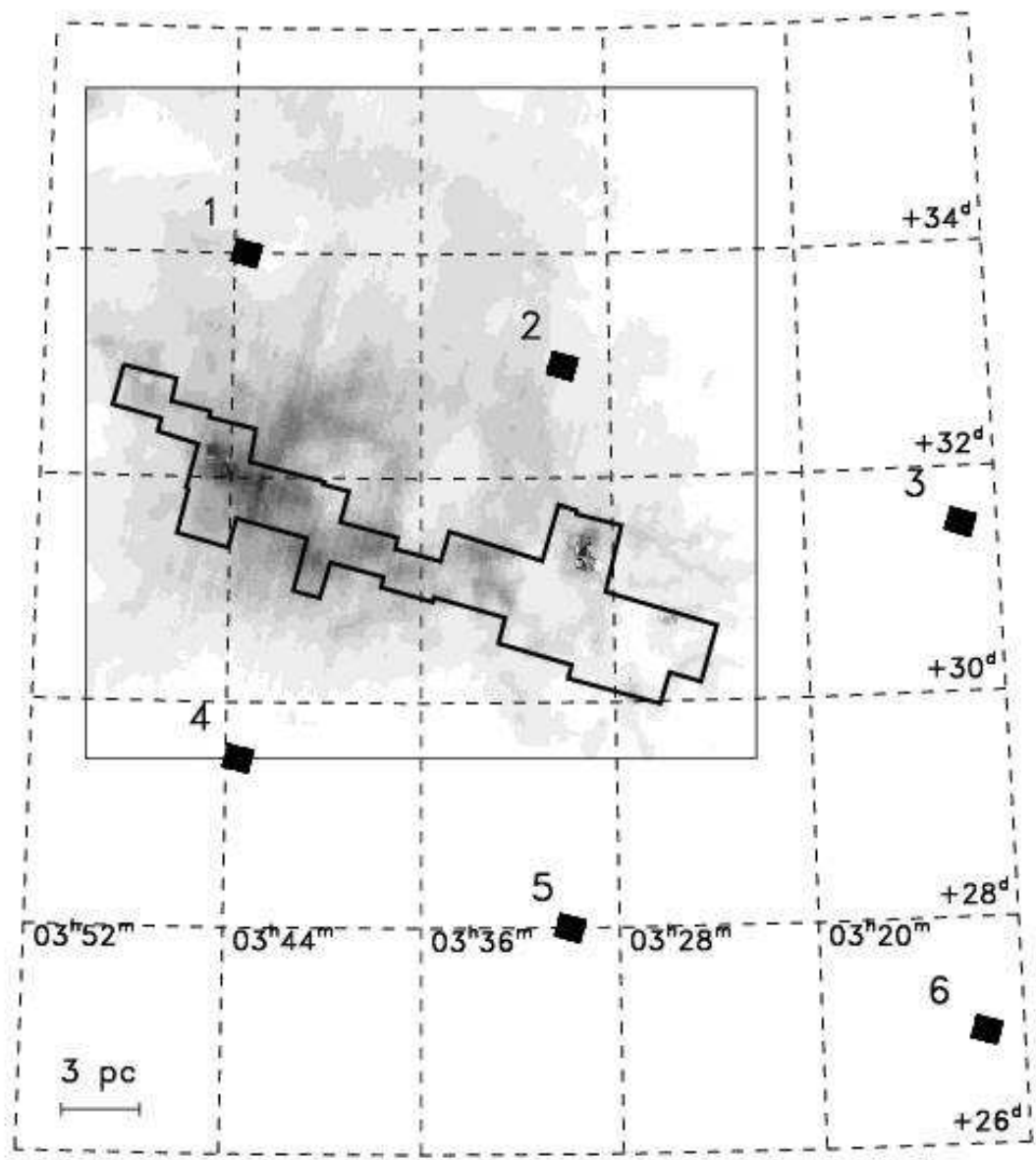}
\caption{Coverage by the c2d IRAC maps. The thick black line indicate
  the region covered by all 4 IRAC bands from which the source
  catalogs are extracted. The off-cloud fields are likewise shown
  (black solid fields) and numbered 1 through 6. The grey scale map is
  the dust column density distribution based on IRAS 60 and 100~$\mu$m
  thermal emission from COMPLETE \citep{schnee05} encompassed by the
  black rectangle. The prominent ring seen in the map of
  \citeauthor{schnee05} is not a true column density feature but
  rather reflects a different, warmer dust population in this region
  \citep{ridge06}.}\label{per_coverage}
\end{figure}

In total, 4.29~degree$^2$ of Perseus was observed with the IRAC camera
in each of 4 bands at 3.6, 4.5, 5.8 and 8.0~$\mu$m on 2004 September 7
and 8. The IRAC camera observes these four bands simultaneously with
one field of view imaged by the 3.6 and 5.8~$\mu$m filters and a
nearby field imaged by the 4.5 and 8.0~$\mu$m filters. Mosaics of our
Perseus observations overlap in all four IRAC bands in a region
covering 3.86~degree$^2$. The mapped area was selected based on the
$^{13}$CO maps of \cite{padoan99b}, including the entire region with
$A_V \gtrsim 2$~mag. Fig.~\ref{per_coverage} shows an overlay of the
observed region on a dust column density map based on IRAS 60 and
100~$\mu$m thermal emission \citep{schnee05}. The cloud was mapped in
two epochs to help in the elimination of transient objects such as
unknown asteroids. The regions around NGC~1333 and IC~348 were mapped
in Guaranteed Time Observations (GTO; \cite{gutermuth06} and
\cite{lada05}, respectively) and observed in one epoch within the c2d
program. Two smaller gaps (per\_gap1, per\_gap2 in Table~\ref{aor})
between the fields mapped by the GTO and c2d were added later and only
observed in one epoch. MIPS observations of the Perseus cloud from c2d
are discussed in a separate paper \citep{rebull06}.

Furthermore 6 off-cloud fields, each of 0.08~degree$^2$, were observed
for statistical comparisons with 0.05~degree$^2$ overlap between all
four IRAC bands. These regions were selected as having a relatively
low extinction ($A_V < 0.5$~mag) from the large scale CO maps by
\cite{dame01} and chosen to cover a range of galactic latitudes around
Perseus. Table~\ref{aor} provides a summary of the observations,
including the distribution of the on- and off-cloud fields.

\clearpage

\begin{table}
\caption{Summary of observations.}\label{aor}
\begin{tabular}{lllll}\hline\hline
Field & Position $(\alpha,\delta)_{2000}$ & Position $(l,b)$ & AOR epoch 1 & AOR epoch 2 \\ \hline
\multicolumn{5}{c}{\emph{AORs mapped in two epochs by c2d}}\\
per\_1    & (03:47:44.0,+32:44:44.0) & $(160.6,-16.9)$ & 0005783552 & 0005790464 \\
per\_2    & (03:46:01.0,+32:29:16.0) & $(160.5,-17.4)$ & 0005783808 & 0005790720 \\
per\_4    & (03:44:53.3,+31:39:00.0) & $(160.9,-18.2)$ & 0005784064 & 0005791232 \\
per\_5    & (03:41:59.0,+31:48:34.0) & $(160.3,-18.4)$ & 0005784320 & 0005791488 \\
per\_6    & (03:40:12.0,+31:26:41.0) & $(160.2,-18.9)$ & 0005784576 & 0005791744 \\
per\_7    & (03:38:27.0,+31:22:11.0) & $(159.0,-19.2)$ & 0005784832 & 0005792000 \\
per\_8    & (03:36:23.0,+31:08:41.0) & $(159.7,-19.7)$ & 0005785088 & 0005792256 \\
per\_9    & (03:33:36.0,+31:08:50.0) & $(159.2,-20.1)$ & 0005785344 & 0005792512 \\
per\_10   & (03:31:00.0,+30:36:36.0) & $(159.1,-20.8)$ & 0005785600 & 0005792768 \\
per\_11   & (03:30:09.0,+31:31:48.0) & $(158.4,-20.2)$ & 0005785856 & 0005793024 \\
per\_13   & (03:28:41.0,+30:34:02.0) & $(158.7,-21.2)$ & 0005786112 & 0005793536 \\
per\_14   & (03:27:12.0,+30:11:49.0) & $(158.7,-21.7)$ & 0005786368 & 0005793792 \\
per\_15   & (03:25:20.0,+30:30:01.0) & $(158.2,-21.6)$ & 0005786624 & 0005794048 \\
per\_16   & (03:27:08.0,+30:39:42.0) & $(158.4,-21.3)$ & 0005786880 & 0005794304 \\
per\_17   & (03:29:39.0,+30:55:28.0) & $(158.7,-20.8)$ & 0005787136 & 0005794560 \\
per\_18   & (03:31:05.0,+31:04:42.0) & $(158.8,-20.4)$ & 0005787392 & 0005794816 \\ \hline
\multicolumn{5}{c}{\emph{AORs mapped in one epoch by GTO team and in one epoch by c2d.}} \\
ic348     & (03:44:21.5,+32:10:16.8) & $(160.4,-17.8)$ & 0003651584 & 0005790976 \\
ngc1333   & (03:29:00.6 +31:18:42.9) & $(158.3,-20.5)$ & 0003652864 & 0005793280 \\ \hline
\multicolumn{5}{c}{\emph{AORs to cover gaps between c2d and GTO areas. Mapped in one epoch by c2d.}} \\
per\_gap1 & (03:29:45.0,+30:54:50.0) & $(158.7,-20.7)$ & 0016034304 & $\ldots$   \\
per\_gap2 & (03:44:36.0,+31:55:38.0) & $(160.7,-17.0)$ & 0016034048 & $\ldots$   \\ \hline
\multicolumn{5}{c}{\emph{Off-cloud regions mapped in two epochs by c2d.}} \\
per\_oc1  & (03:43:30.0,+34:00:00.0) & $(159.1,-16.5)$ & 0005795072 & 0005798912 \\
per\_oc2  & (03:30:00.0,+33:00:00.0) & $(157.4,-19.1)$ & 0005795328 & 0005799168 \\
per\_oc3  & (03:13:30.0,+31:30:00.0) & $(155.3,-22.3)$ & 0005795584 & 0005799424 \\
per\_oc4  & (03:43:30.0,+29:30:00.0) & $(162.1,-20.0)$ & 0005795840 & 0005799680 \\
per\_oc5  & (03:30:00.0,+28:00:00.0) & $(160.7,-23.0)$ & 0005796096 & 0005799936 \\
per\_oc6  & (03:13:30.0,+27:00:00.0) & $(158.1,-26.0)$ & 0005796352 & 0005800192 \\ \hline
\end{tabular}
\end{table}

\clearpage

\section{Data reduction}
The data reduction started from the IRAC basic calibrated data (BCD)
images supplied by the Spitzer Science Center (SSC) from their S11.4.0
pipeline. The SSC pipeline BCD data are delivered with dark and bias
levels subtracted and flat field corrected. Instrumental signatures,
bad pixels or pixels saturated by bright sources are furthermore
masked. Finally the BCD images are flux calibrated into physical units
(MJy~sr$^{-1}$). To arrive at the final source catalogs our data
reduction continues from the BCD images by 1) correcting additional
background and image effects, 2) creating a mosaic from individual
frames, 3) extracting sources from the final mosaic and 4) merging the
catalogs across bands. A detailed description of the processing is
described in the documentation for the delivery of c2d data
\citep{delivery3} and further discussed in \cite{harvey06serpens}.

In the first step, the c2d pipeline improves the BCD images by
additional masking of bad pixels not picked up by the SSC pipeline
reduction, correcting muxbleed and column pull-down defects occurring
near moderately bright sources and finally correcting the so-called
``first frame'' effect for the band 3 (5.8~$\mu$m) images (see
\cite{harvey06serpens} for further discussion of these
corrections). In the second step, mosaics are created from the frames
from observations in the two different epochs and from short exposure
(0.4 second) data. The combination of the two epoch observations
eliminates transient effects including most asteroids in the
mosaics. The inclusion of the short exposure data allows masking of
likely saturated areas in the longer integration frames.

As a third step, sources are extracted using a c2d developed tool,
\textit{c2dphot} \citep{harvey06c2dphot}. \textit{c2dphot} extracts
sources using digitized source profiles and calculates photometry for
each source including uncertainty in the derived fluxes. The source
extractor recognizes ``small'' extended sources (i.e., sources better
fit by a two-axis ellipsoid than a point source profile) and provides
an estimate of the size of these. These extended sources
\citep[sources with ``image type'' 2 in the delivered c2d
  catalogs][]{delivery3} are included in our general catalog with
fluxes extracted from aperture photometry. However, we leave them out
of the color-color/color-magnitude diagrams and discussions in
Sect.~\ref{yso_pop} and onwards in this paper for consistent
photometry of sources in our high quality catalogs (i.e., when
comparing unresolved and extended sources). The accuracy of the flux
calibration for the sources is estimated to $\approx$~15\%. The
extraction is applied to each of the single epoch datasets together
with the combined mosaics. A uniform flux cut-off of 0.05~mJy is
applied across the regions and bands. The source lists from each
dataset are cross-checked serving as an additional test of transient
detections.

In the fourth step, the lists are merged by cross-identifying sources
from each of the IRAC bands followed by merging with the MIPS
24~$\mu$m \citep{rebull06} and 2MASS catalogs. In the band merging an
IRAC source is accepted as identified in two bands if the positional
agreement is better than 2.0$''$. Cross identifications with MIPS
24~$\mu$m require positional agreement better than 4.0$''$ and with
2MASS better than 2.0$''$.

\section{Results}
\subsection{Description of the overall cloud environment}
\begin{figure}
\plotone{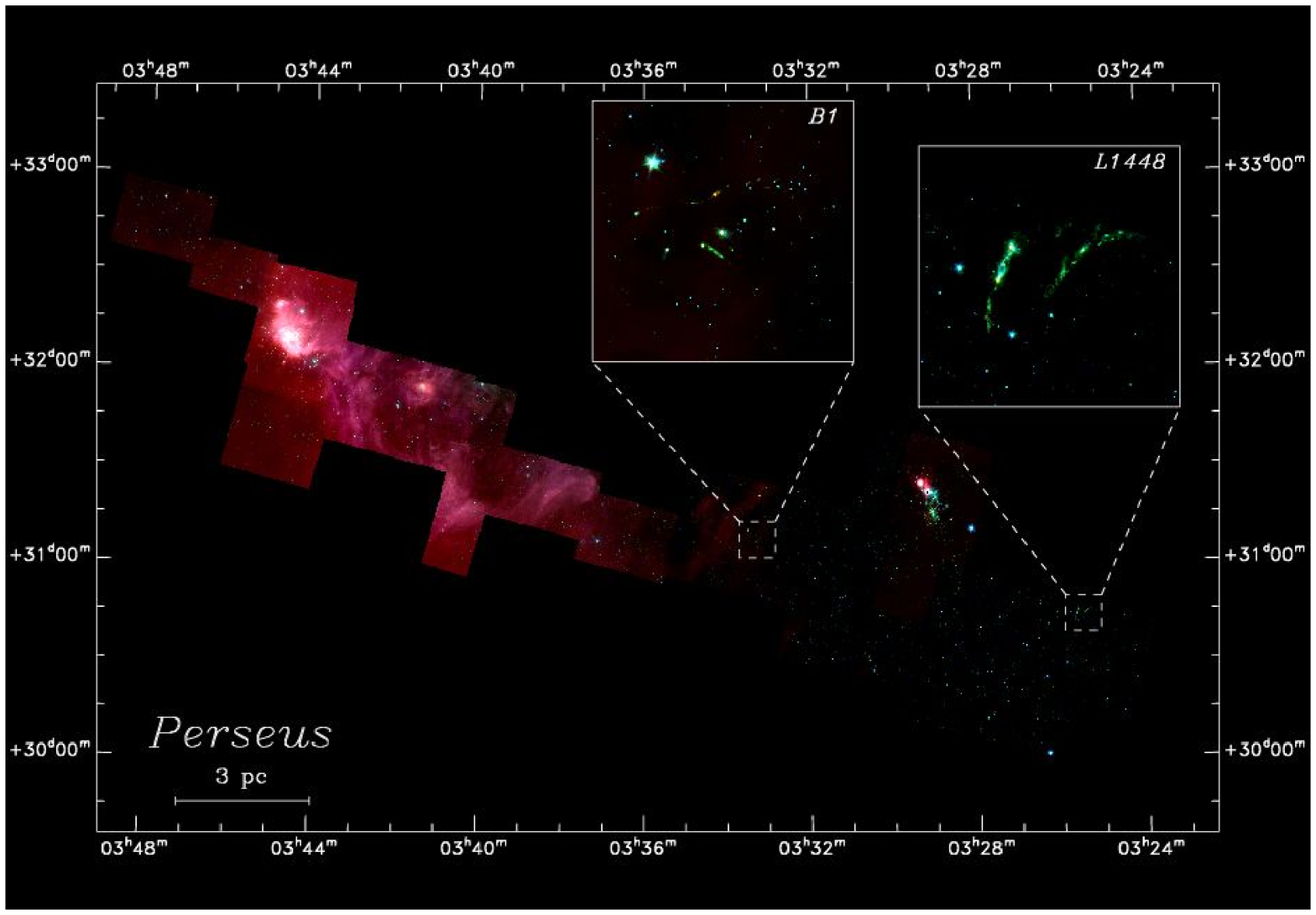}
\caption{Three color image of the Perseus region with IRAC1
  (3.6~$\mu$m; blue), IRAC2 (4.5~$\mu$m; green) and IRAC4 (8.0~$\mu$m;
  red).}\label{color_rgb}
\end{figure}

A three color composite of the Perseus region is presented in
Fig.~\ref{color_rgb}, showing IRAC bands 1 (IRAC1; 3.6~$\mu$m), 2
(IRAC2; 4.5~$\mu$m) and 4 (IRAC4; 8.0~$\mu$m). For display purposes
the image in this figure does not include the short integration data
which otherwise are used for photometry of saturated sources. For
comparison, the extinction contours from \cite{enoch05} are overlaid
on the band 2 image in Fig. 3, and prominent regions
identified. Fig.~\ref{first_blackwhite} shows greyscale images
(logarithmic stretch) of the cloud for each of the IRAC bands. As
pointed out above, the field observed by the IRAC1 and IRAC3 bands is
offset from the field observed by the IRAC2 and IRAC4 bands which is
clearly seen by comparing the panels of
Fig.~\ref{first_blackwhite}. Only the overlap region is shown in
Fig.~\ref{color_rgb} and used in the subsequent analysis. The emission
seen in the IRAC bands highlight different features of the cloud: band
1 shows the largest fraction of stars, band 2 is, in particular,
sensitive to H$_2$ rotational transitions arising in shocked gas
associated with the outflows, whereas bands 3 and 4 includes emission
from the polycyclic aromatic hydrocarbon (PAH) features at 6.2 and
7.7~$\mu$m.

Besides the two main clusters, NGC~1333 and IC~348, a striking feature
in the image is the Perseus ring (sometimes referred to as
G159.6-18.5) in the eastern part of the cloud, which is seen in all
bands - most prominent in the 8.0~$\mu$m (likely PAH) emission. It has
been suggested to be a supernova remnant \citep{fiedler94} or an
\ion{H}{2} region associated with a late O/early B star, HD~278942
\citep{dezeeuw99,andersson00}. \cite{ridge06} argue that the shell is
located behind the cloud (although it may be ``touching'' it).

\begin{figure}
\plotone{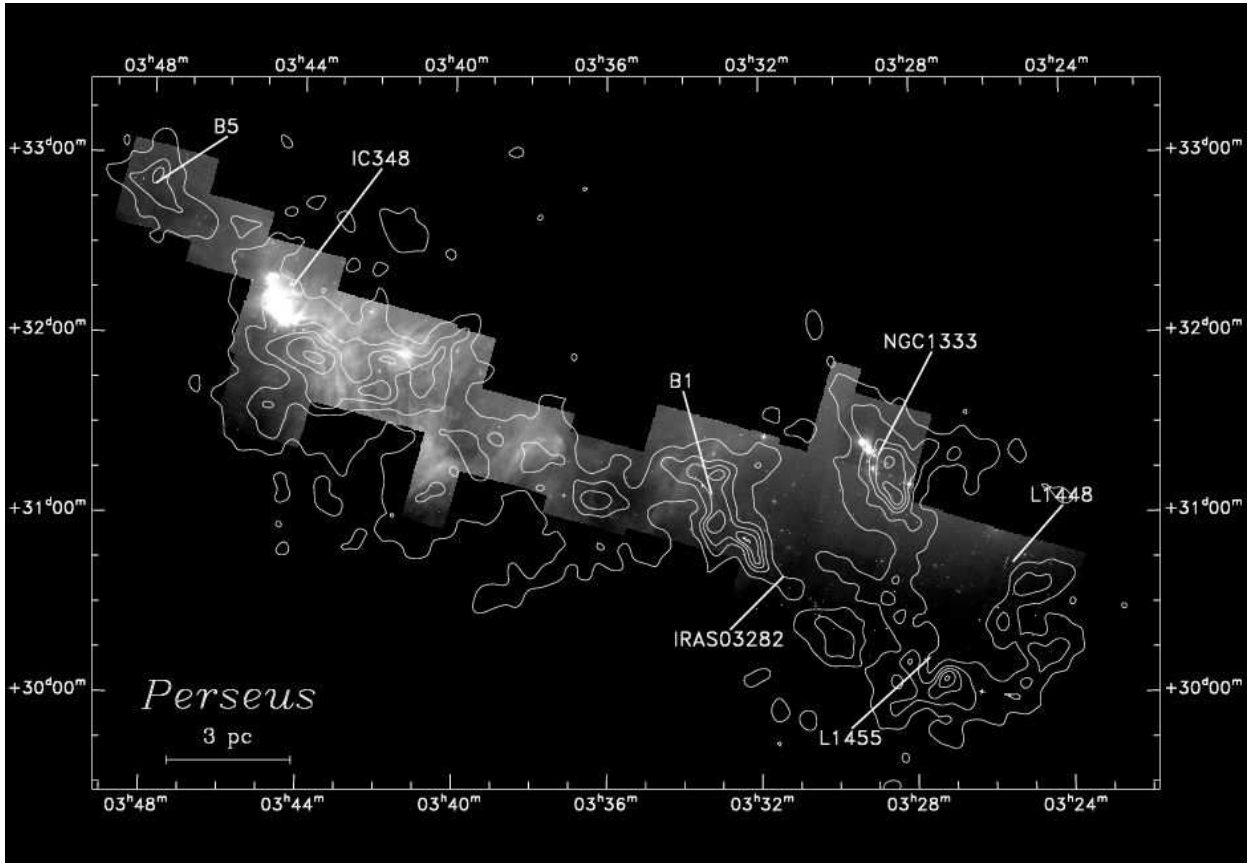}
\caption{IRAC2 image with prominent clusters and cores indicated. The
  dashed contours indicate an extinction map with levels corresponding
  to $A_V=2,4,6,8,10$~mag.}\label{overview}
\end{figure}

\begin{figure}
\plotone{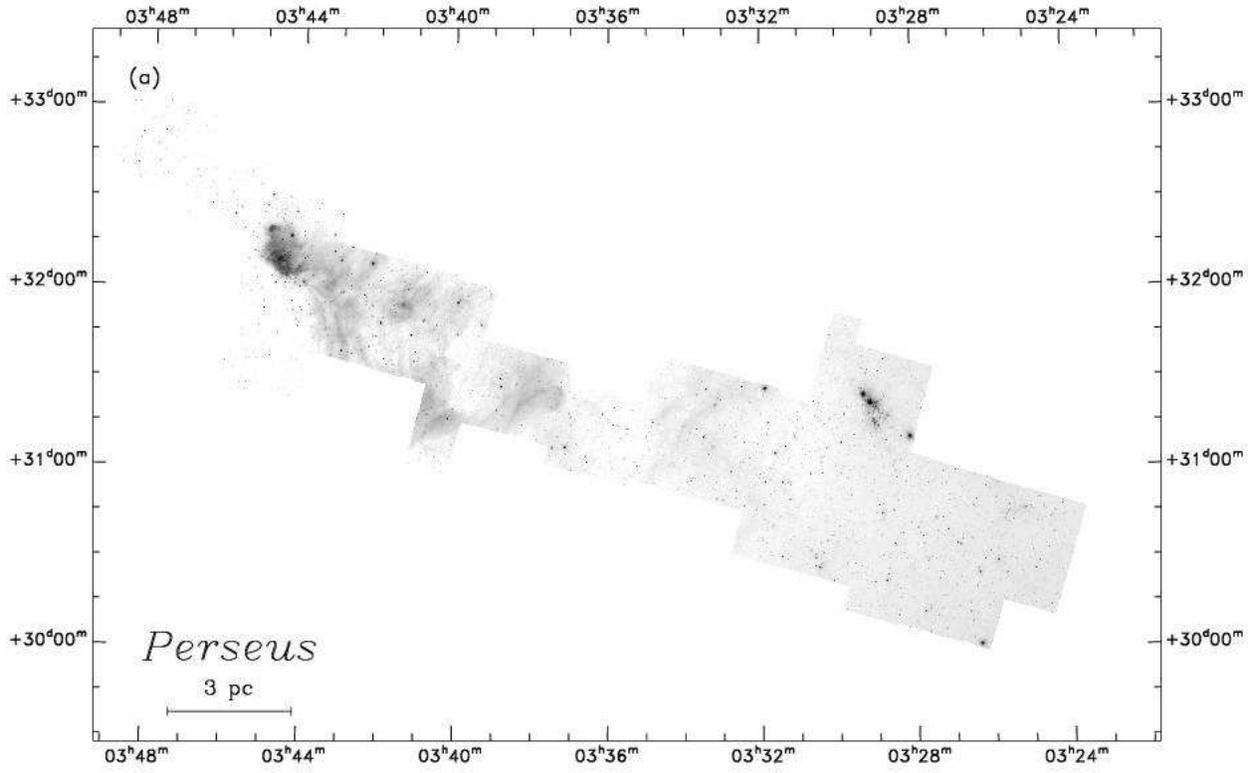}
\caption{Black/white maps of the entire region (shown with a
  logarithmic stretch) in \textit{(a)} IRAC band 1 (3.6~$\mu$m),
  \textit{(b)} band 2 (4.5~$\mu$m), \textit{(c)} band 3 (5.8~$\mu$m),
  and \textit{(d)} band 4 (8.0~$\mu$m).}\label{first_blackwhite}
\end{figure}

The western part of the cloud shows less PAH emission than the eastern
part, but harbors a number of prominent outflows, in particular in the
B1, L1448 and L1455 regions (see inserts). We will return to a
discussion of some of these outflows in Sect.~\ref{outflows}. The
extinction map (Fig.~\ref{overview}) demonstrates, however, that
regions of medium-high column density are not isolated to a particular
part of the cloud. In contrast, the Serpens cloud exhibits smaller and
more isolated regions of high extinction, where also the highest
number of Class I YSOs are found \citep{harvey06serpens}. In Perseus,
regions of high extinction are likewise found close to IC~348 and
NGC~1333 but these are in no way unique. There is also no clear
correlation between the regions of high extinction and the extended
emission seen in the IRAC images.

\begin{figure}
\plotone{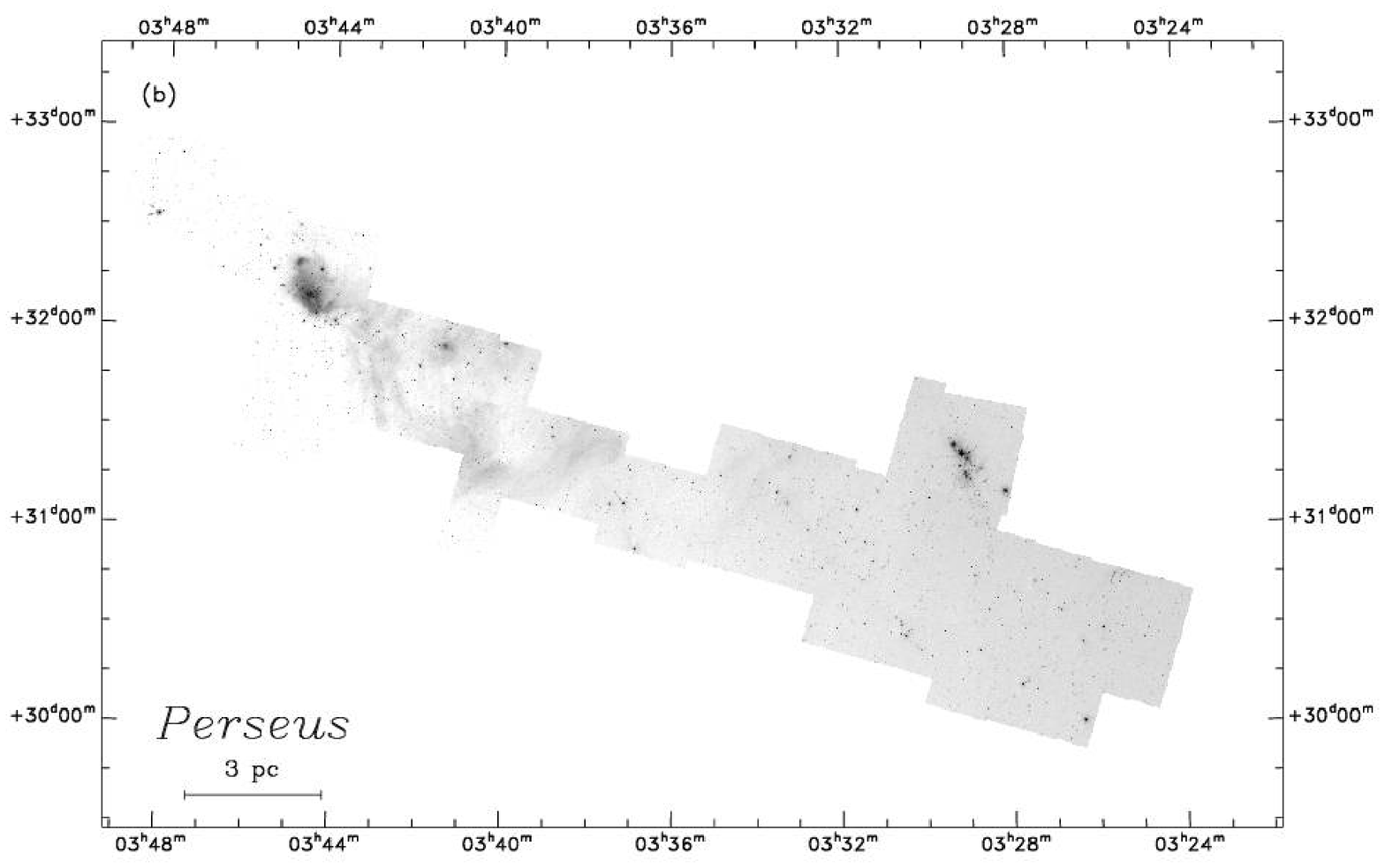}
\end{figure}
\begin{figure}
\plotone{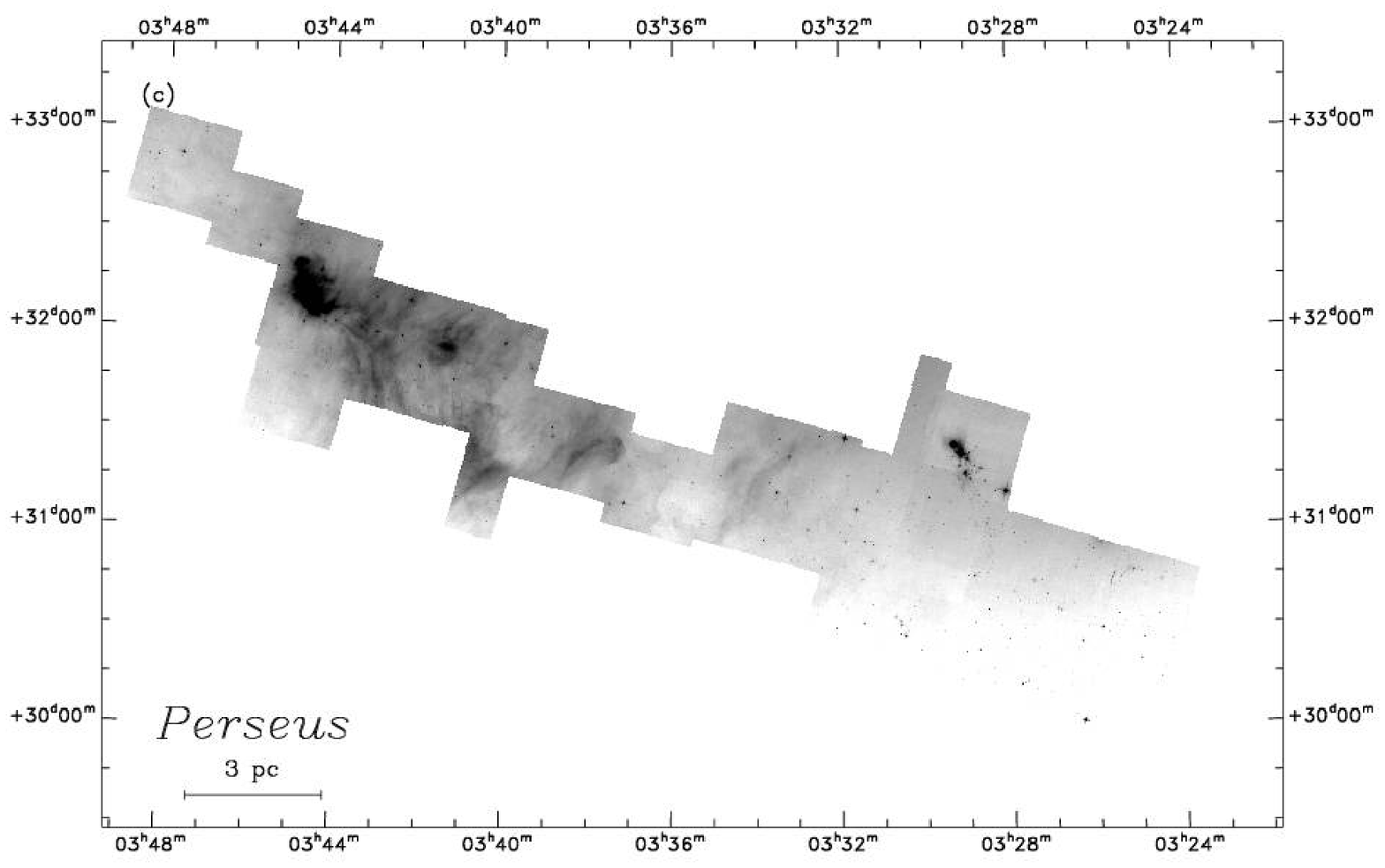}
\end{figure}
\begin{figure}
\plotone{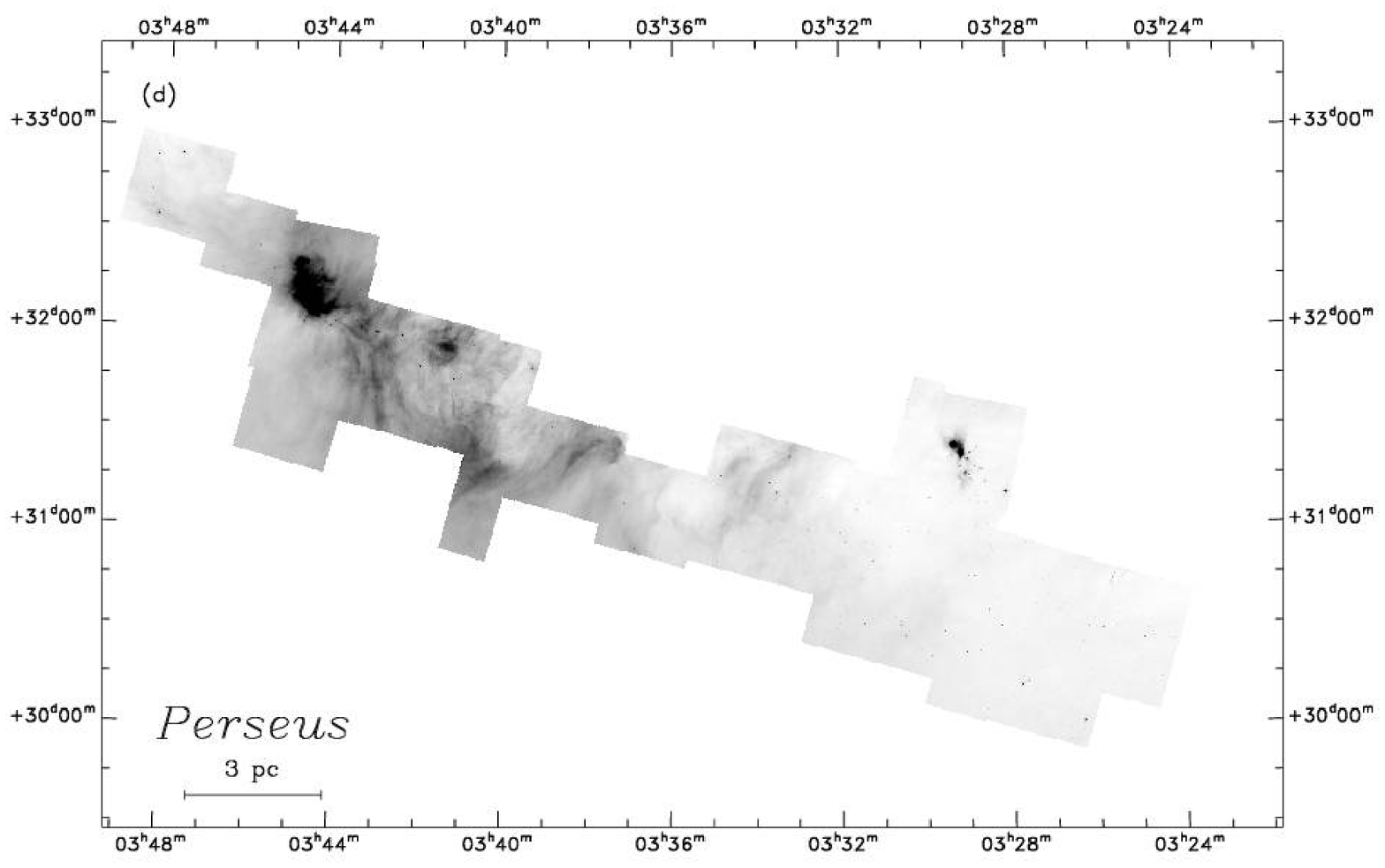}
\end{figure}

\subsection{Catalogs and statistics}\label{highqualitydef}
In total 123327 sources were detected by the Spitzer observations in
at least one IRAC band. Table~\ref{detect1} summarizes the number of
sources detected in the Perseus maps with S/N ratios of at least
7. Sources which have multiple counterparts within the radius used in
the bandmerging are excluded from this ``high quality'' catalog, which
we use in the subsequent sections. This corresponds to selecting all
sources with detections of quality ``A'' or ``B'' in one of the IRAC
bands from the delivered c2d catalogs \citep{delivery3}. Of all the
sources in the catalog the largest fraction is most likely background
stars picked up in the most sensitive 3.6 and 4.5~$\mu$m bands. This
is clearly illustrated in Table~\ref{detect2} where the number of
detected sources are split up in the different IRAC bands: with
70,000--110,000 sources detected in IRAC bands 1 and 2 and
10,000--15,000 in band 3 and 4. Of all the reliable detections, about
6700 sources were detected in all four IRAC bands with about 20\% of
these not previously detected by 2MASS. Conversely, 65 of the 7933
2MASS sources (0.8\%) included in the Perseus mosaic were not detected
in any IRAC band. Most of these 65 sources are located at the edge of
the Spitzer maps explaining why they are not included in the
catalog. Finally about 541 of the four band detections are found to be
extended in one or more bands (classified with ``image type'' 2 in the
c2d catalogs). These are left out of the color-color and
color-magnitude diagrams in the following sections and our high
quality catalog therefore consists of 6192 four band detections.

\clearpage

\begin{table}
\caption{Detection of sources with S/N $\ge 7\sigma$ toward Perseus
  (total numbers). }\label{detect1}
\begin{tabular}{ll}\hline\hline
Detection in at least one IRAC band               & 123327 \\
Detection in all 4 IRAC bands                     &   6733 \\
Detection in 3 IRAC bands                         &   8438 \\
Detection in 2 IRAC bands                         &  50925 \\
Detection in 1 IRAC band                          &  57231 \\[1.0ex]
Detection in 2MASS only$^{a}$                     &     65 \\
Detection in IRAC only                            & 114975 \\
Detection in 4 IRAC bands and not 2MASS$^{a}$     &   1380 \\\hline
\emph{Excluding extended sources:} & \\
Four band detections                              & 6192 \\
Four band detections with 2MASS association$^{a}$ & 5115 \\
Detected in IRAC1+2 and 2MASS$^{a}$               & 7487 \\ \hline
\end{tabular}

$^{a}$A source is counted as detected by 2MASS if it has a S/N of at
least 10 in both $H$ and $K_s$.
\end{table}

\clearpage

\begin{table}
\caption{Detection of sources toward Perseus (per band).}\label{detect2}
\begin{tabular}{lllll} \hline\hline
                                                      & 3.6~$\mu$m     & 4.5~$\mu$m     & 5.8~$\mu$m     & 8.0~$\mu$m     \\ \hline
Detections with                                       &                &                &                &          \\
$\ldots$S/N of at least 7                             & 110585         &    76425       &    14999       &     9318 \\
$\ldots$S/N of at least 10                            &  85913         &    60042       &    10189       &     7010 \\
$\ldots$S/N of at least 15                            &  53592         &    34619       &     6644       &     4810 \\
Final sample (excluding extended sources)             & 104341         &    74877       &    14832       &     8997 \\
$\ldots$2MASS ass. (S/N at least 10 in $H$ and $K_s$) &   7532         &     7778       &     7206       &     5434 \\\hline
\end{tabular}
\end{table}

\clearpage

\begin{figure}
\resizebox{\hsize}{!}{\includegraphics{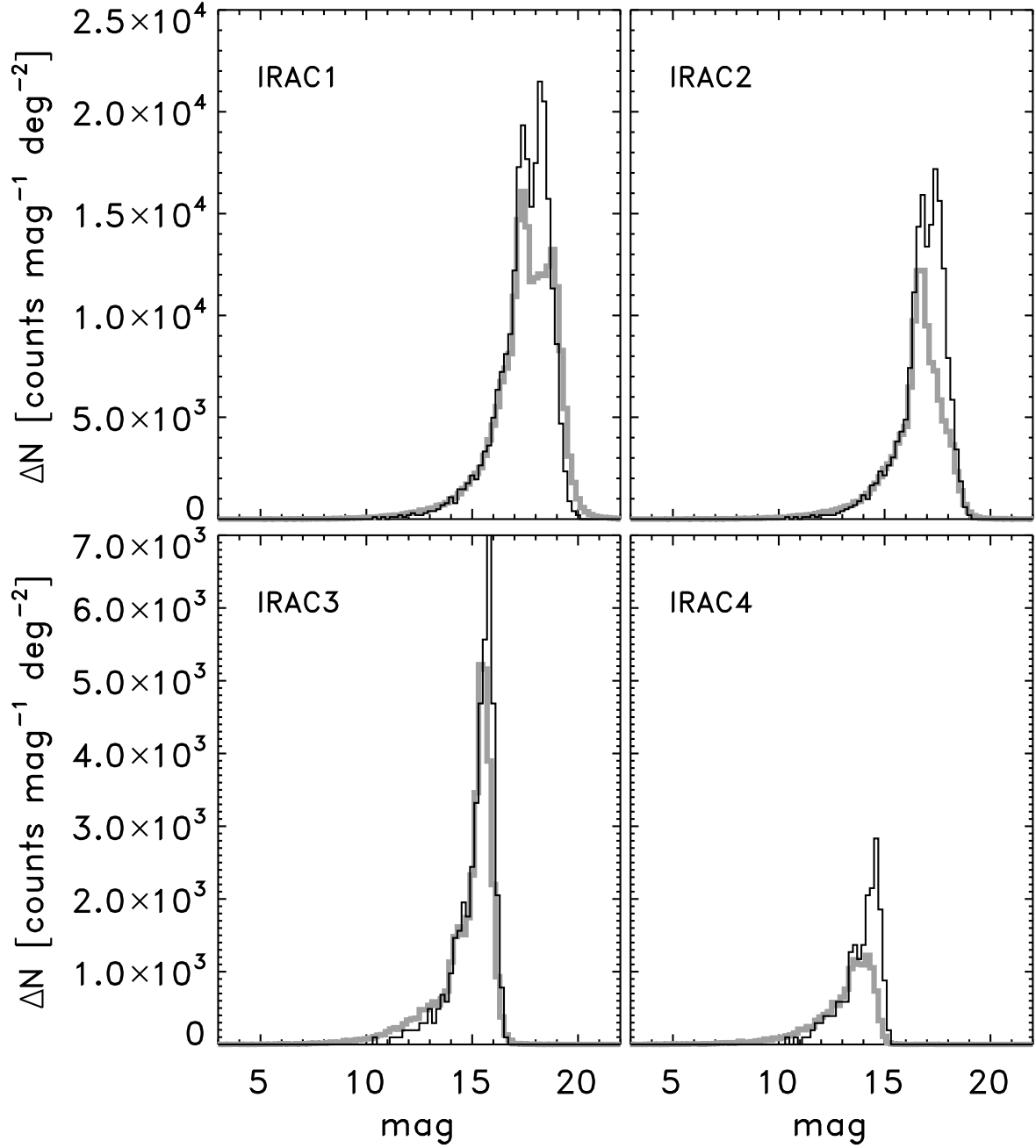}}
\caption{Differential source counts for the on- and off-cloud regions
  (grey and black, respectively).}\label{diffcount}
\end{figure}

Fig.~\ref{diffcount} shows the differential source counts for the
cloud and ``off-cloud'' areas. The turn over at faint magnitudes
illustrates the completeness limit for each IRAC band: 17.5 mag for
IRAC1, 16.5 mag for IRAC2, 14.5 mag for IRAC3 and 13.5 mag for
IRAC4. An interesting feature is clearly seen in the comparison
between the on- and off-cloud fields: the populations in the two agree
well in the bright tail of the distribution, but in the faint end
there is a clear excess of sources in the off-cloud fields.

\clearpage

\begin{figure}
\resizebox{\hsize}{!}{\includegraphics{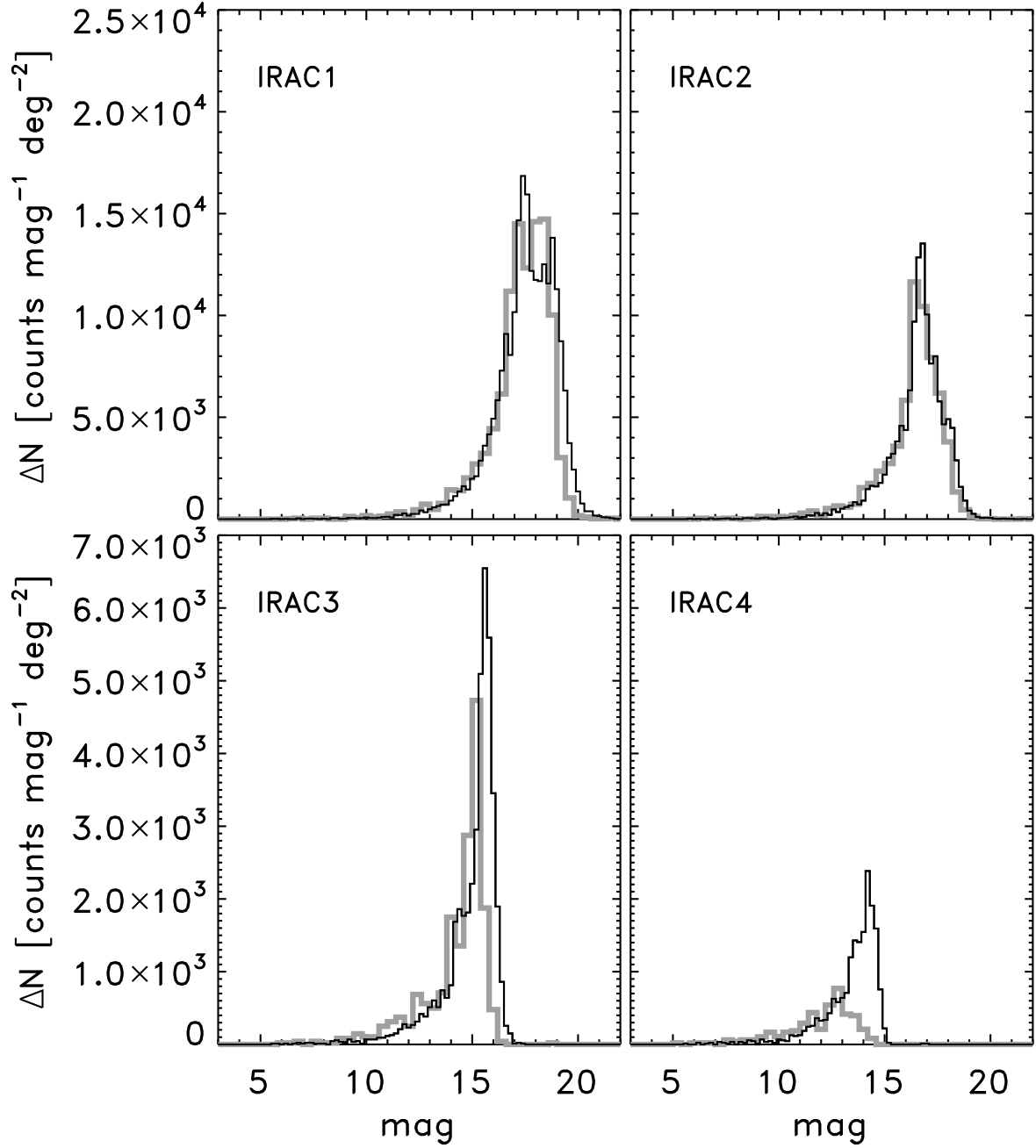}}
\caption{Differential source counts for a low and high extinction
  field (black and grey lines, respectively).}\label{diffcount_av}
\end{figure}

Naturally some difference could be expected between the on- and
off-cloud fields: the off-cloud fields are selected to have low
extinction whereas most of the on-cloud area has $A_V \gtrsim
2$~mag. Fig.~\ref{diffcount_av} compares the differential source
counts from a high extinction ($A_V \gtrsim 6$~mag) and a low
extinction ($A_V \lesssim 2$~mag) field based on the extinction map in
Fig.~\ref{overview}. This figure shows that there is an excess of
faint sources in the low extinction field, similar to that seen in the
off-cloud fields, relative to the counts in the high extinction
fields, i.e., extinction in the cloud shifts the differential source
count distribution toward fainter magnitudes.

The double peak seen most clearly in the distribution of IRAC band 1
sources at 17--18 mag. in both the on- and off-cloud fields appears to
be real. It is not an artifact of the calibration or combination of
the different mosaics or frames (e.g., long and short exposures); it
is not associated, e.g., with sources only detected in one epoch; and
it is seen across the cloud and not related to any specific region. It
therefore seems to reflect either the distribution of stars or
background galaxies toward Perseus, but based on this dataset it is
not possible to say which is more likely. Deep observations and a
detailed comparison to a larger sample of fields could possibly shed
further light on this issue.

\begin{figure}
\resizebox{0.9\hsize}{!}{\includegraphics{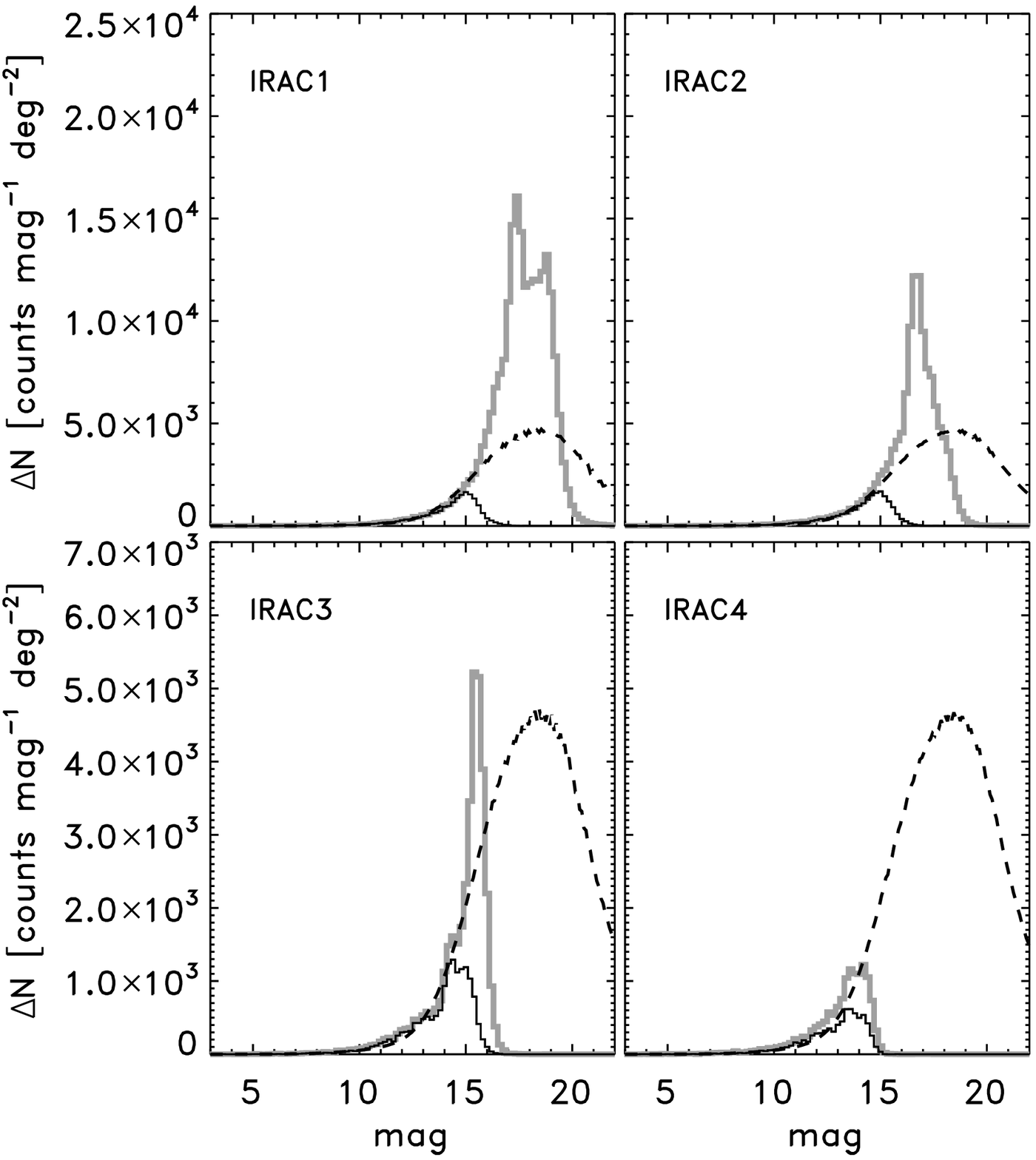}}
\caption{Differential source counts for all on-cloud sources as in
  Fig.~\ref{diffcount} (grey) and objects which can be classified as
  stars through fits to their 2MASS and IRAC measurements by a
  (reddened) photosphere (black). Also shown are the predictions for
  the star counts from the models of \cite{wainscoat92} (black dashed
  line). An excess of background galaxies is clearly seen at faint
  magnitudes from the comparison between the distribution of stars
  from the model predictions and the actual number counts. Note that
  the fainter stars in the Spitzer observations are not classified as
  such due to lack of 2MASS data. The difference between the black
  solid and dashed lines at faint magnitudes therefore reflect the
  completeness of the 2MASS data rather than an actual discrepancy
  between the IRAC observations and the model
  predictions.}\label{diffcount_star}
\end{figure}

Finally Fig.~\ref{diffcount_star} compares the on-cloud differential
source counts to the predicted galactic star counts from the models of
\cite{wainscoat92} calculated using the C-version of the models by
J. Carpenter. The model predictions agree well with the observed star
counts at bright magnitudes, but underestimate the source counts at
brightnesses fainter than 13--15 mag. The reason is likely
contributions from background galaxies which start to dominate the
counts there. For this comparison and our subsequent analysis, the
stars are separated from the remaining sources by fitting a (possibly
reddened) photosphere \citep{lai05} to the 2MASS and IRAC
measurements. This method relies on 2MASS measurements and does not
properly identify fainter stars only picked up in the IRAC bands. It
is seen that the observed population of stars agree well with the
predictions of the \cite{wainscoat92} models but also that the
background galaxies outnumber the predicted star counts by more than
3:1 at faint magnitudes in some of the bands. This effect is naturally
more pronounced for these data than corresponding Serpens data
\citep{harvey06serpens} since Perseus is located further from the
Galactic plane and the stellar density therefore is lower than in
Serpens.

\section{The YSO population}\label{yso_pop}
\subsection{Colors of the sources}
With more than 6000 sources in the final high quality catalog detected
in all four IRAC bands it is of great interest to establish general
criteria for selecting samples of candidate young stellar objects. A
number of such schemes have been suggested based on IRAC colors alone
or in combination with 2MASS or MIPS colors
\citep[e.g.,][]{allen04,megeath04,gutermuth04,muzerolle04}. With the
Spitzer observations, a high number of photometric data points are
obtained that provide strong constraints on the loci in the
color-color and color-magnitude planes characterizing young stellar
objects.

\begin{figure}
\resizebox{\hsize}{!}{\includegraphics{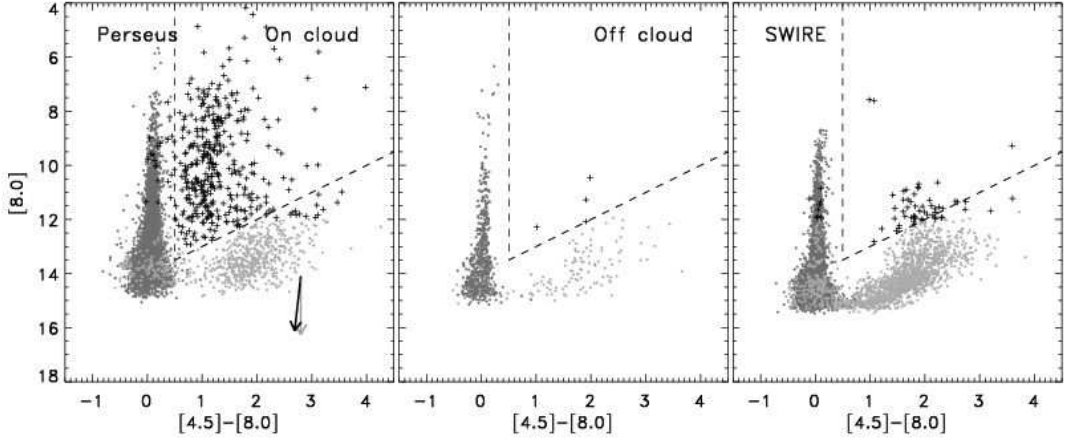}}
\caption{$[8.0]$ vs. $[4.5]-[8.0]$ color-magnitude diagrams based on
  the high quality catalogs (see text) with the Perseus ``on-cloud''
  regions (left), ``off-cloud'' region (middle) and trimmed SWIRE
  catalog (right). In each panel stars have been indicated by dark
  grey dots, YSO candidates by black plus signs and other sources by
  light grey dots. The distinction between YSO candidates and other
  sources was predominantly made based on this diagram with YSO
  candidates taken from the upper right part of the diagram above and
  right of the two dashed lines with a few additional sources added
  based on their $[8.0]-[24]$ colors (see discussion in text). Two
  extinction vectors are shown corresponding to $A_K$ = 5 mag. The
  gray vector was derived for diffuse ISM regions by
  \cite{indebetouw05}.  The black vector, appropriate for the dense
  regions found in molecular clouds and cores, was derived from deep
  near-infrared and Spitzer observations of dense cores
  \citep{huard06}.}\label{i24_v4}\label{firstcolor}
\end{figure}

\begin{figure}
\resizebox{\hsize}{!}{\includegraphics{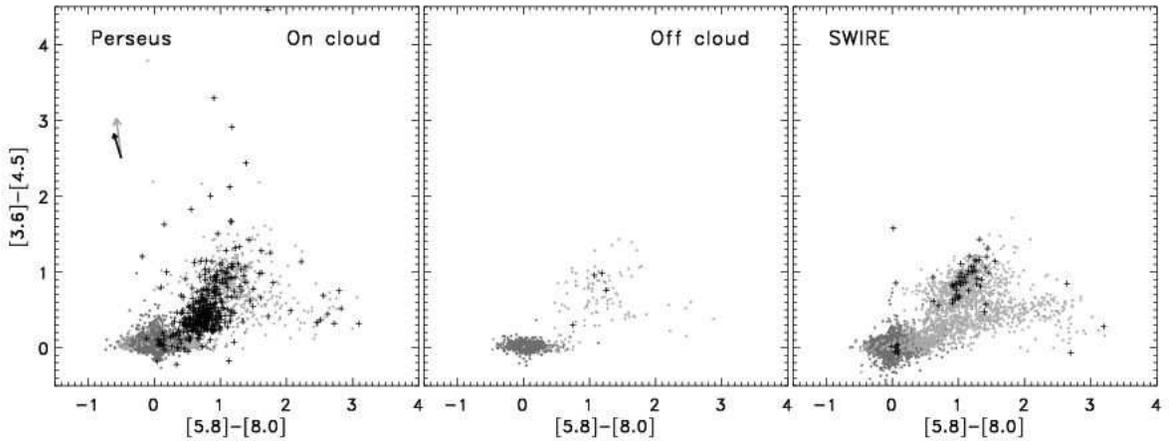}}
\caption{$[3.6]-[4.5]$ vs. $[5.8]-[8.0]$ color-color diagram for the
  ``on-cloud'' (left), ``off-cloud'' (middle) and SWIRE (right)
  fields. Sources, symbols and extinction vectors as in
  Fig.~\ref{i24_v4}.}\label{i34_v12}
\end{figure}

\begin{figure}
\resizebox{\hsize}{!}{\includegraphics{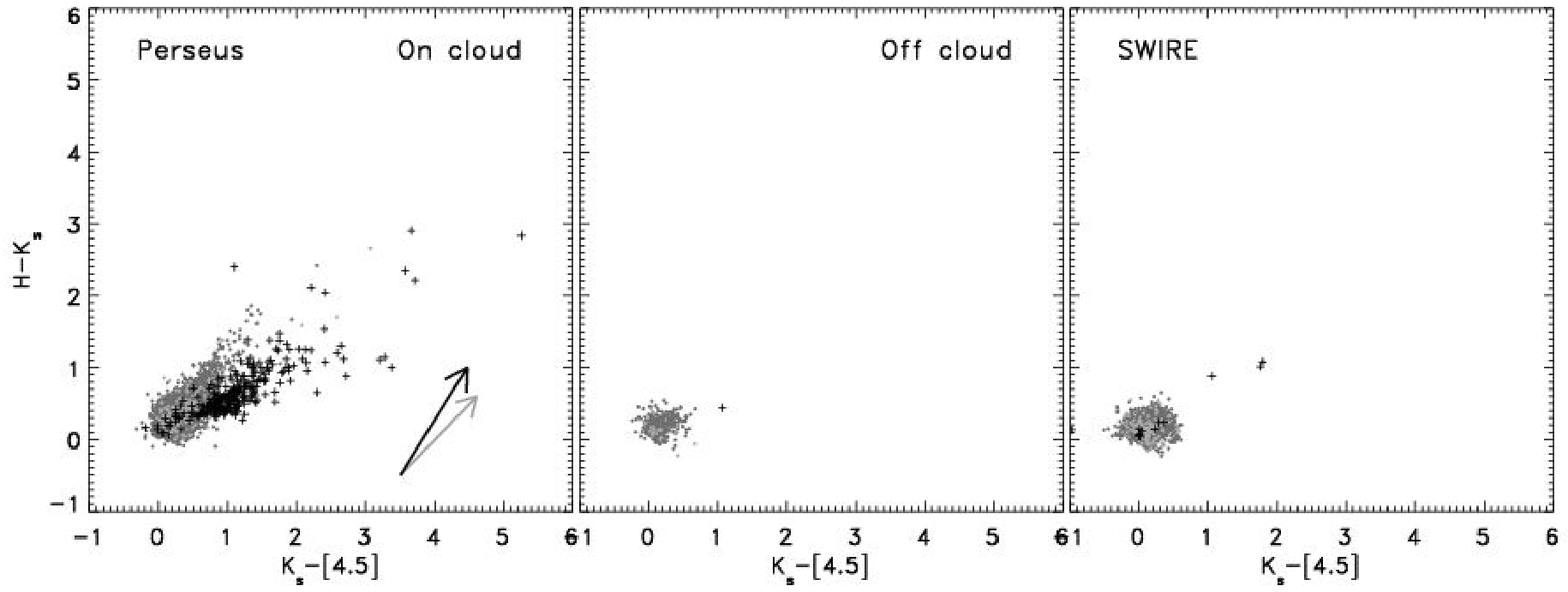}}
\caption{As in Fig.~\ref{firstcolor} and \ref{i34_v12} but for a
  $H$-$K_s$ vs. $K_s$-[4.5] color-color diagram. All sources with high
  quality detections in the IRAC bands 1 and 2 (see text) and
  10$\sigma$ detections in $H$ and $K_s$ included. Otherwise sources
  and symbols as in Fig.~\ref{i24_v4} and \ref{i34_v12}; extinction
  vectors shown are for $A_K=2$~mag.}\label{ihk_vk2}\label{lastcolor}
\end{figure}

A first task is to separate the YSO population from the remainder of
the sources toward the cloud, ``contaminating the sample'', mainly
stars and background galaxies. Figs.~\ref{firstcolor}-\ref{lastcolor}
show color-magnitude and color-color diagrams for the Perseus cloud
compared to the off-cloud and catalogs and data from the SWIRE Elais
N1 data set \citep{surace04}, which is used to estimate the
extragalactic component of our catalogs. The SWIRE catalog was
processed exactly like the cloud datasets resulting in a catalog
consisting of 103557 sources distributed over 6.3~degree$^2$. For a
better comparison to the Perseus data the SWIRE catalog was trimmed
down to the same area as covered by the Perseus observations by
randomly selecting sources. The SWIRE data also go somewhat deeper due
to longer integration time, so the catalog of SWIRE sources was
furthermore trimmed to a similar completeness limit as the Perseus
data. In the analysis only sources with high quality detections (as
described in Sect.~\ref{highqualitydef}) have been included. In each
of the diagrams the group of stars constitute a distinct group
centered around zero colors. Comparing the $[8.0]$ vs. $[4.5]-[8.0]$
color-magnitude diagram for the Perseus on-cloud region to the
off-cloud regions and SWIRE data there obviously is a significant
excess of sources with $[4.5]-[8.0] > 0.5$ and $[8.0]$ magnitudes
brighter than $14-([4.5]-[8.0])$ as indicated by the dashed lines in
the top right panel. In total, 400 such objects are found that as a
first cut represent the candidate YSOs toward Perseus. In the
``on-cloud'' and, in particular, the SWIRE diagrams a large population
of fainter red objects is seen, which are likely dominated by
background galaxies. At faint magnitudes the diagram becomes
inadequate for separating these three basic group of objects and it
becomes necessary to look at the full photometric data for
classification. In particular, stars are separated from the remaining
objects by fitting a (possibly reddened) photosphere \citep{lai05} as
described above.

Some sources do not show an excess in the IRAC bands but do show an
excess in the MIPS 24~$\mu$m observations discussed by
\cite{rebull06}. The combination of the long wavelength IRAC and MIPS
observations are expected to be particularly sensitive to infrared
excesses since photospheric colors are expected to be zero and has
previously been applied to the selection of young stellar objects
\citep{muzerolle04}. As above, sources with excesses in the MIPS band
can be selected from $[24]-[8]>0.7$ and $[24] < 12-([24]-[8.0])$. This
selection adds another 37 candidate YSOs to our catalogs. These
include objects with faint excesses only detectable at longer
wavelengths (i.e., more evolved ``Class III'' objects). Embedded YSOs,
with some degree of PAH emission contributing to the IRAC band 4
emission (shifting them rightward in Fig.~\ref{i24_v4} to the faint
side of the magnitude cut-off) or with certain geometries of their
circumstellar envelopes, can also fall into this category,
however. Still, this is a small fraction ($\approx$~9\%) compared to
our total list of 400 YSO candidates but illustrates how the inclusion
of even longer wavelength data adds to the information about the
sample of YSOs.

A small fraction, 3--4\%, of the non-star objects in the SWIRE and
off-cloud diagrams are bright enough at 8 or 24~$\mu$m to be
misclassified as YSOs. If the same fraction applies to the Perseus
diagram, about 24 ($\approx$~6\%) of the YSOs in the on-cloud region
might in fact be extragalactic objects. This number is an upper limit,
though: any extragalactic object will be behind the cloud, extincted,
and thus moved in the direction of the extinction vector toward the
faint side of the magnitude cut-off (non-vertical, dashed) line in the
$[8.0]$ vs. $[4.5]-[8.0]$ diagram in Fig.~\ref{i24_v4}.

Fig.~\ref{i34_v12} and Fig.~\ref{ihk_vk2} show the location of the
different groups of sources in two other diagrams that have been used
for classifying YSOs from samples of Spitzer sources, the
$[3.6]-[4.5]$ vs. $[5.8]-[8.0]$ \citep{megeath04} and $H-K_s$
vs. $K_s-[4.5]$ \citep{gutermuth04} color-color diagrams. The
locations of the YSO candidates agree well with the predicted
locations in these two diagrams: in the $[3.6]-[4.5]$
vs. $[5.8]-[8.0]$ diagram, the YSOs are expected to fall in a group
with colors around $([3.6]-[4.5],[5.8]-[8.0])=(0.7,0.5)$ or redder
according to \cite{allen04} and \cite{megeath04}. It is also clear,
however, that this diagram in itself is not sufficient to separate the
YSO candidates from, i.p., the background galaxy populations. In
contrast, the $H-K_s$ vs. $K_s-[4.5]$ diagram appears to be better
suited for selecting YSO candidates. The detected galaxies and stars
are largely overlapping with colors close to zero (compare, e.g., to
the SWIRE diagram) whereas the YSO candidates have redder $H-[4.5]$
colors, in particular. The drawback to this diagram is the requirement
that the YSOs are all detected at $H$ and $K_s$. It turns out that
only 280 (70\%) of the YSO candidates have 2MASS associations, i.e.,
10$\sigma$ detections in both $H$ and $K_s$ (305 or 76\% of the YSO
candidates are associated with 2MASS sources with 5$\sigma$ detections
in $H$ and $K_s$). The sources not detected by 2MASS are likely more
deeply embedded sources or brown dwarfs and therefore the selection of
YSO candidates using 2MASS criteria is somewhat biased against such
objects. One could, on the other hand, suspect that the $H-K_s$
vs. $K_s-[4.5]$ diagram would identify sources not included in the
other diagrams, given that only detections in IRAC bands 1 and 2 are
required. This, however, seems not to be the case as only a few
``other'' sources are identified in the red part of the diagram - as
also shown by the comparison to the off-cloud diagram. The best way to
separate faint YSOs from galaxies seems to be with much deeper JHK
data.

\subsection{Classification of YSO candidates}\label{yso_cc}
With a sample of candidate YSOs selected, a next natural step is to
search for trends in evolution between these. The most used
classification scheme for YSOs is based on their spectral slope,
$\alpha=\frac{{\rm d}\log \lambda F_\lambda}{{\rm d}\log\lambda}$
\citep{lada87}, dividing the YSOs into Class I, II and III with
increasingly blue SEDs. This classification scheme is thought to
represent (roughly) the evolution of YSOs from their embedded
protostellar phases through their pre-main sequence stages. The
\citeauthor{lada87} classification scheme has subsequently been
expanded with a group of ``flat spectrum'' YSOs suggested to represent
a transitional phase between the Class I and II stages
\citep{andre94,greene94}. Earlier in the evolution of low-mass
protostars, the so-called ``Class 0'' objects \citep{andre93}, are
thought to represent the most deeply embedded YSOs. This class of
objects is defined by the SEDs at submillimeter wavelengths where they
emit more than 0.5\% of of their total bolometric luminosity at
wavelengths longer than 350~$\mu$m and not by their mid-infrared
signatures: previously these objects were in general not detected at
mid-infrared wavelengths. A number of Class 0 objects are known in
Perseus and we will return to these in Sect.~\ref{class0}. In the
remainder of this section, however, we will not distinguish these
deeply embedded protostars from the Class I objects.

\begin{figure}
\resizebox{\hsize}{!}{\includegraphics{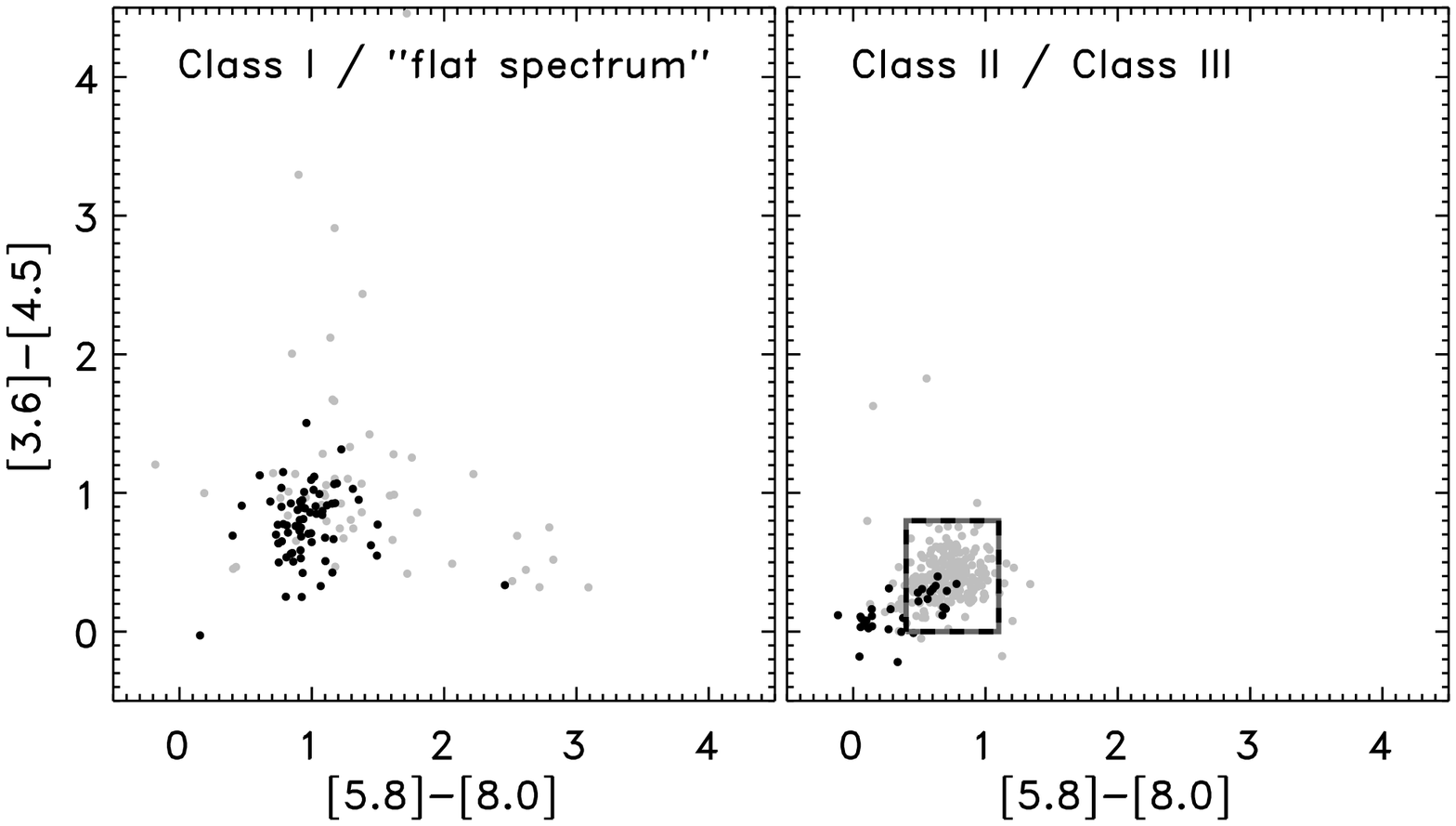}}
\caption{$[3.6]-[4.5]$ vs. $[5.8]-[8.0]$ color-color diagrams for the
  YSO sample. The symbols are colored according to their
  classification based on their spectral slope $\alpha$
  ($\alpha=\frac{{\rm d}\log \lambda F_\lambda}{{\rm d}\log\lambda}$)
  in the scheme of \cite{lada87}: Class I objects ($\alpha \ge 0.3$)
  and ``flat spectrum'' ($-0.3 \le \alpha < 0.3$) sources are shown in
  the left panel with grey and black symbols, respectively. Likewise
  Class II ($-1.6 \le \alpha < -0.3$) and Class III objects ($\alpha <
  -1.6$) are shown in the right panels with grey and black symbols,
  respectively. The box delineating Class II objects from
  \cite{allen04} has furthermore been indicated in the right panel
  (black/grey dashed rectangle).}\label{lada_cc}
\end{figure}

For each of the sources in the Perseus YSO catalog, a spectral index
is assigned based on its photometry from 2MASS $K_s$ through MIPS 24
\micron.  Fig.~\ref{lada_cc} plots the sources with different spectral
indexes in the $[3.6]-[4.5]$ vs. $[5.8]-[8.0]$ diagrams following the
division by \cite{greene94} where Class I objects have $\alpha \ge
0.3$, ``flat spectrum'' objects have $-0.3 \le \alpha < 0.3$, Class II
objects $-1.6 \le \alpha < -0.3$ and Class III objects $\alpha <
-1.6$. With this scheme 243 of the YSOs in Perseus are classified as
Class II, 71 as ``flat spectrum'' and 54 as Class I YSOs.

Based on a theoretical study, \cite{allen04} suggested that the Class
II objects, i.e., pre-main sequence stars with disks, would have
colors $0.0< [3.6]-[4.5]<0.8$ and $0.4<[5.8]-[8.0]<1.1$ whereas the
Class I objects, i.e., objects still embedded in a circumstellar
envelope, would have colors $[3.6]-[4.5]>0.8$ and/or
$[5.8]-[8.0]>1.1$. This division between Class I and II objects is
seen to be in good agreement with the $\alpha$ classification
described above for 219 (90\%) of the Class II and 50 (93\%) of the
Class I objects. 42 (59\%) of the ``flat spectrum'' sources fall in
the region of Class I objects in the scheme of \cite{allen04} and 28
(39\%) in the region of Class II objects underscoring that this group
lies between the Class I and II objects. It may not be that surprising
that these trends are found, given that the estimate of $\alpha$ and
the color diagrams essentially are based on the same measurements. The
classification is also not unambiguous: Class II objects with edge-on
disks for example have colors similar to the Class I objects.

A cross-check with the SIMBAD astronomical database reveals that 217
(or 54\%) of the YSO candidates from the list in this paper are
associated with known objects - almost all YSO-like - within a search
radius of 5$''$. Expanding the search radius to 10$''$ only increases
this number slightly to 228. Interestingly only 35\% and 39\% of the
Class I and ``flat spectrum'' sources are associated with known
sources, whereas the numbers for the Class II and III objects are 60\%
and 72\%, respectively. It therefore appears that the new YSO
candidates in our catalog are predominantly faint, embedded sources
with significant mid-infrared excesses, as should be expected from the
high sensitivity of Spitzer compared to previous surveys included in
SIMBAD - in particular those based on IRAS data.

Our selection of YSO candidates does not properly identify Class III
objects: of our YSO candidates, 32 objects have $\alpha < -1.6$
consistent with Class III objects (16 of these 32 are selected as YSO
candidates on basis of the criterion involving their IRAC4 and
MIPS~24~$\mu$m colors, a high fraction compared to the only
$\approx$~9\% of the entire YSO population selected in this way).
Given that the emission from such objects is largely photospheric and
therefore difficult to distinguish from stars, this sample is not
complete. Additional data from other wavelengths are necessary to
address these issues and properly cross-identifying YSOs
(Lai~et~al.~in prep.). One of the goals for c2d is to include such
data - also to refine the classification schemes and thus our
understanding of the evolution of YSOs.

\subsection{Spatial distribution of YSOs}
\begin{figure}
\epsscale{.9}
\plotone{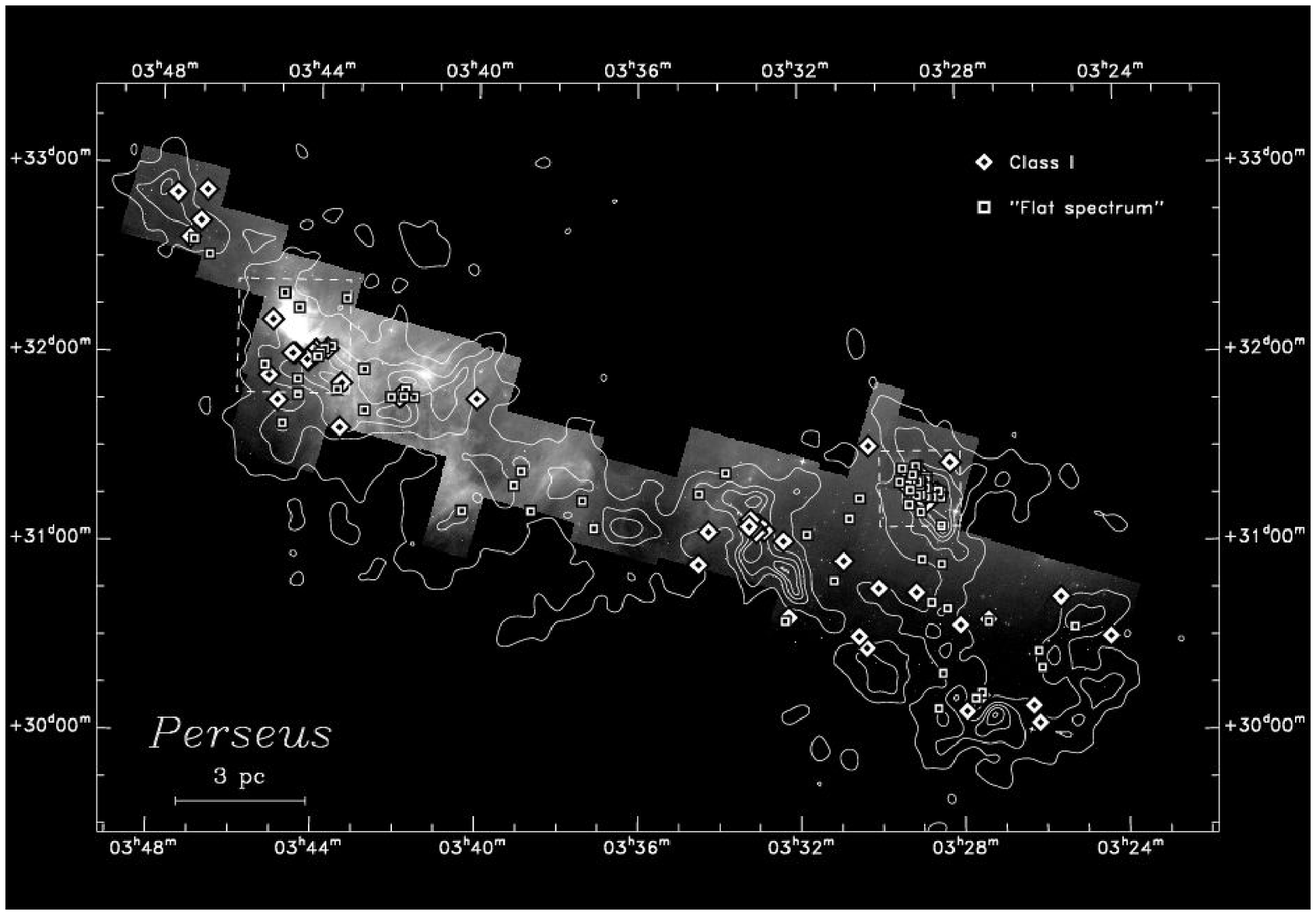}
\plotone{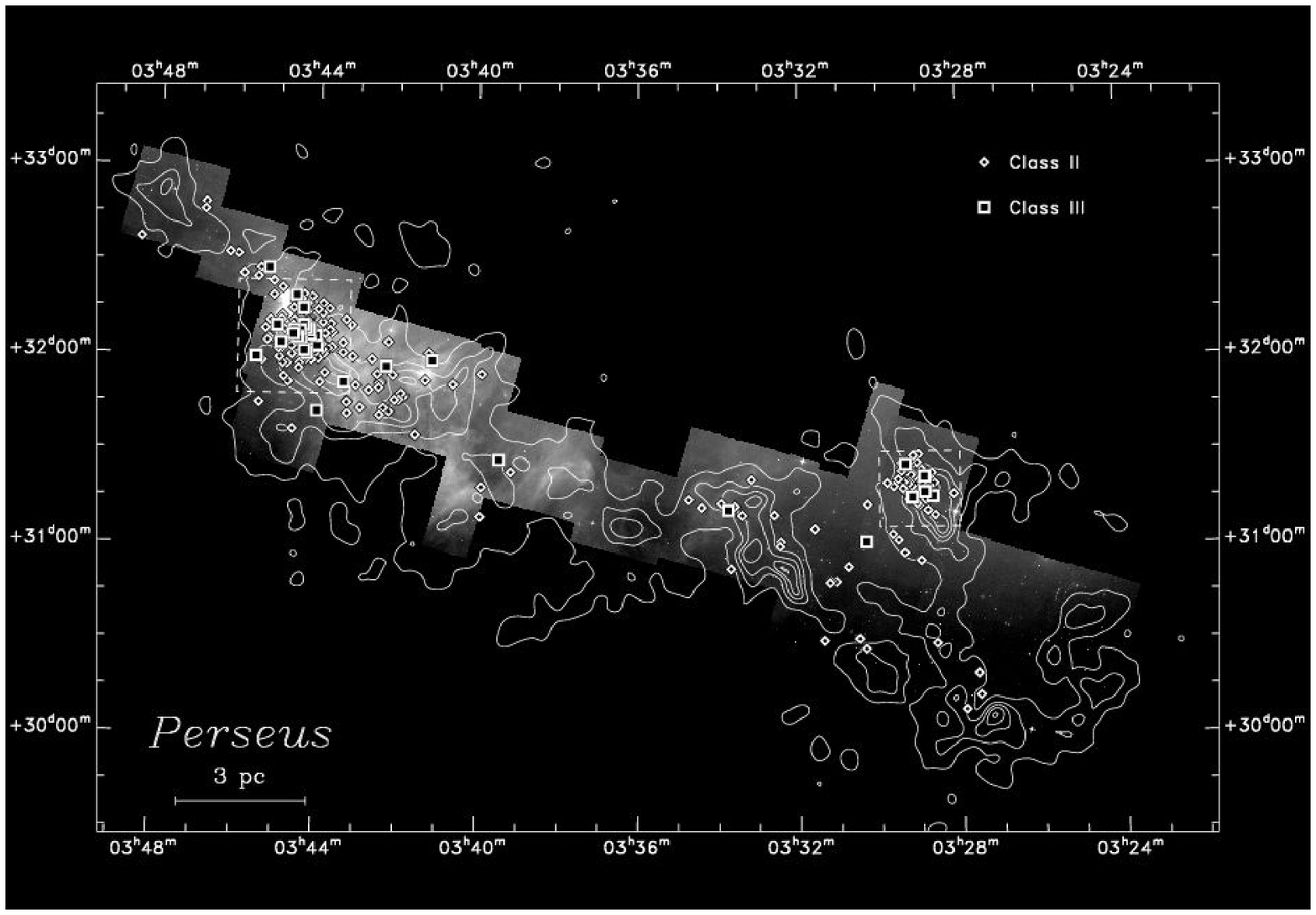}
\epsscale{1}
\caption{Distribution of Class I objects (upper) and Class II objects
  (lower) plotted on top on a 4.5~$\mu$m image of Perseus with an
  extinction map overlay (see also Fig.~\ref{overview}). The adopted
  boundaries for IC~348 and NGC~1333 have been indicated by the dashed
  lines.}\label{source_dist}
\end{figure}

\begin{figure}
\resizebox{\hsize}{!}{\includegraphics{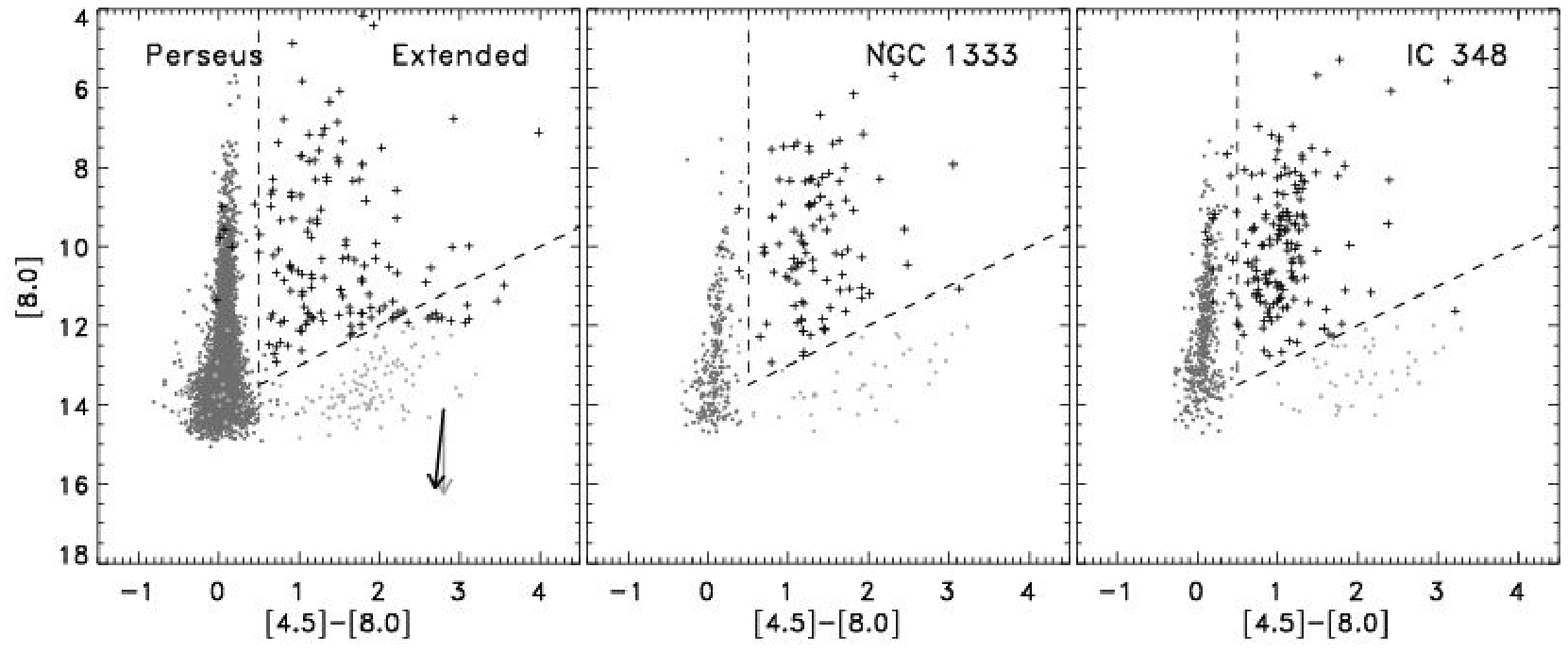}}
\resizebox{\hsize}{!}{\includegraphics{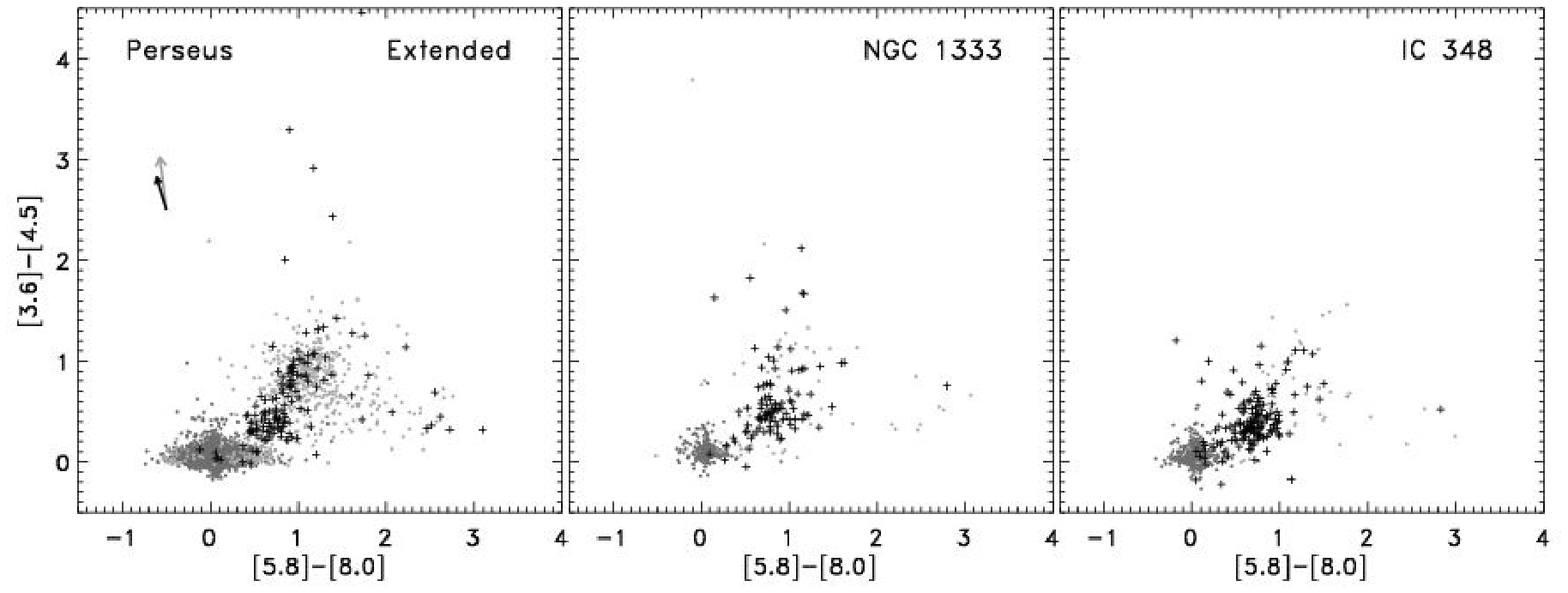}}
\resizebox{\hsize}{!}{\includegraphics{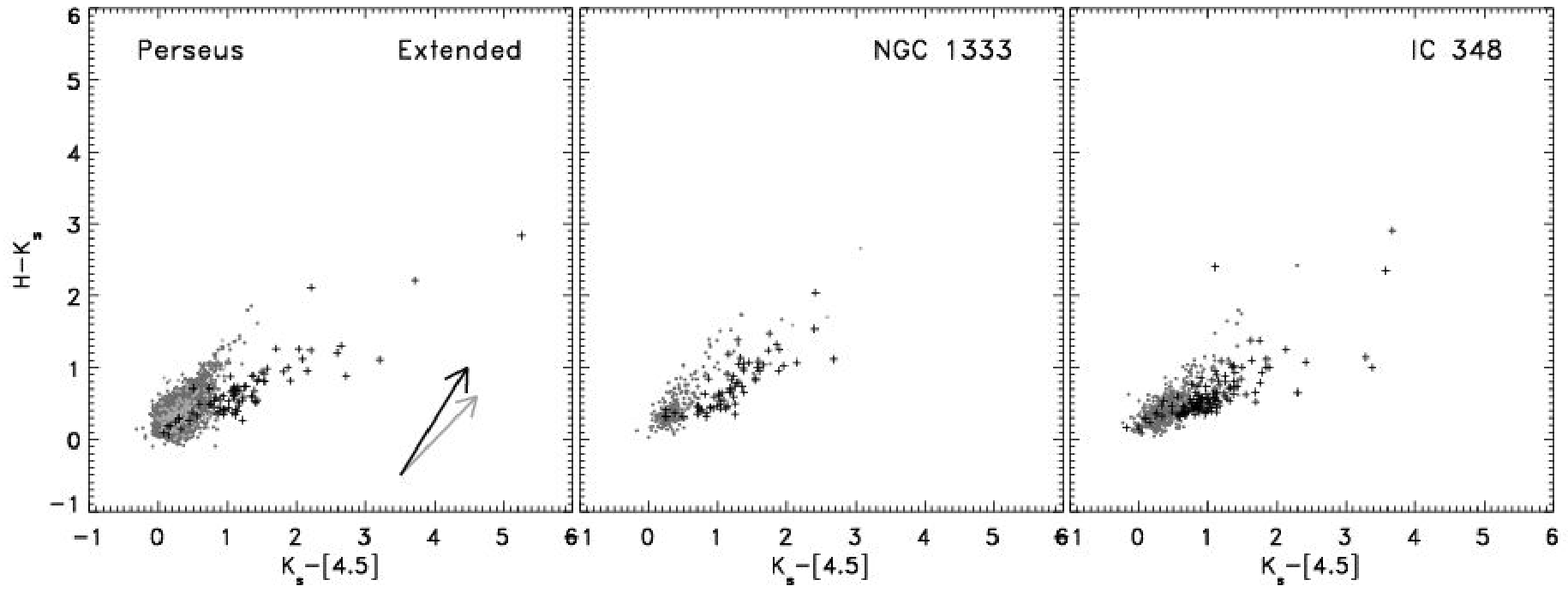}}
\caption{Color-color and color-magnitude diagrams from the extended
  cloud excluding NGC~1333 and IC~348 (left), NGC~1333 only (middle)
  and IC~348 only (right). As in Fig.~\ref{firstcolor}-\ref{lastcolor}
  stars have been indicated by dark grey dots, YSO candidates by black
  plus signs and other sources by light grey
  dots.}\label{color_n1333_ic348}
\end{figure}

Fig.~\ref{source_dist} shows the identified Class I and II YSO
candidates plotted over the 4.5~$\mu$m IRAC image. The two clusters
around NGC~1333 and IC~348 clearly stand out with a high density of
sources. Fig.~\ref{color_n1333_ic348} shows the color-color and
color-magnitude diagrams for the extended cloud (excluding the
clusters) compared to each of these two clusters. The NGC~1333 region
was defined as ranging from 03:28:00 to 03:30:00 in Right Ascension
and +31:06:00 to +31:30:00 in Declination while the IC~348 region was
defined as ranging from 03:43:12 to 03:46:00 in Right Ascension and
+31:48:00 to +32:24:00 in Declination (all J2000) including the areas
studied in the GTO programs. The numbers of total YSO candidates and
numbers of Class I and II objects in each of these regions are
summarized in Table~\ref{cluster_diff}. Significant differences exist
between IC~348, NGC~1333 and the remaining cloud. The two clusters
contain similar numbers of YSO candidates and contain about half of
the total YSO candidates in the surveyed area. The spatial density of
YSO candidates is 5--6 times higher in each of the clusters compared
to the extended cloud. In the extended cloud most of the YSOs appear
to be following the extinction with concentrations of predominantly
Class I objects around B1, B5, L1448 and L1455 and an additional
number of Class II objects about 40\arcmin\ southwest of the main
concentration of sources in IC~348. \cite{cambresy06} recently
suggested the existence of a cluster separate from the main IC~348
cluster at this location through maps of the spatial density of
sources derived from 2MASS data. This cluster is not included in the
statistics for IC~348 below but rather together with the remaining
cloud.

Table~\ref{cluster_diff} also shows that there is a significant
difference in the relative numbers of Class I/''flat spectrum'' and
Class II objects among the two clusters and the extended cloud with
the more embedded Class I objects being more numerous in NGC~1333
compared to IC~348 whose YSO population predominantly consist of Class
II YSOs. A difference between the two clusters is consistent with
near-infrared studies suggesting an age of about 2~Myr for IC~348
\citep{luhman03} and $<1$~Myr for NGC~1333
\citep{lada96,wilking04}. The relative number of Class I and ``flat
spectrum'' objects compared to the overall YSO candidates is
significantly higher in the extended cloud compared to these two
clusters, however, suggesting that a significant fraction of the
recent and current star formation in Perseus is going on in the
extended cloud including the smaller groups surrounding B1, L1455 and
L1448. In fact 61\% of the Class I YSOs toward Perseus are located
outside the two main clusters compared to only 28\% of the Class II
YSOs. Even within the group of Class I and ``flat spectrum'' sources
it seems that the extended cloud contains more of the red Class I
objects compared to the numbers in NGC~1333. A more detailed study of
the clustering properties of the different groups of YSOs in these
regions - and comparison to other similar selections of YSOs in other
cloud regions - could shed further light on differences in YSOs formed
in large clusters compared to objects formed in small groups or in
relative isolation.

\clearpage

\begin{table}
\caption{Number of YSO candidates in IC~348 and NGC~1333 compared to the remaining cloud. Numbers in parentheses indicate the relative fraction of YSOs of different classes compared to the total number of YSOs in the specific regions.}\label{cluster_diff}
\begin{tabular}{lllll}\hline\hline
   & Total surveyed area & IC~348 & NGC~1333 & Remaining cloud \\ \hline
Area                                 & 3.86~deg$^2$ & 0.36~deg$^2$& 0.17~deg$^2$& 3.33~deg$^2$ \\
YSO candidates                       & 400         & 158          & 98          & 144         \\
YSOs per sq. degree                  & 103.6       & 438.9        & 576.5       & 43.2        \\
Class I                              & 54 (14\%)   & 11 (7\%)     & 10 (10\%)   & 33 (23\%)   \\
``Flat spectrum''                    & 71 (18\%)   & 11 (7\%)     & 25 (26\%)   & 35 (24\%)   \\
Class II                             & 243 (61\%)  & 116 (73\%)   & 58 (59\%)   & 69 (48\%)   \\
Class III                            & 32 (8\%)    & 20  (13\%)   &  5 (5\%)    &  7 (5\%)    \\ \hline
\end{tabular}
\end{table}

\clearpage

\section{Deeply embedded protostars and associated outflows}\label{outflows}\label{class0}
\begin{figure}
\resizebox{\hsize}{!}{\includegraphics{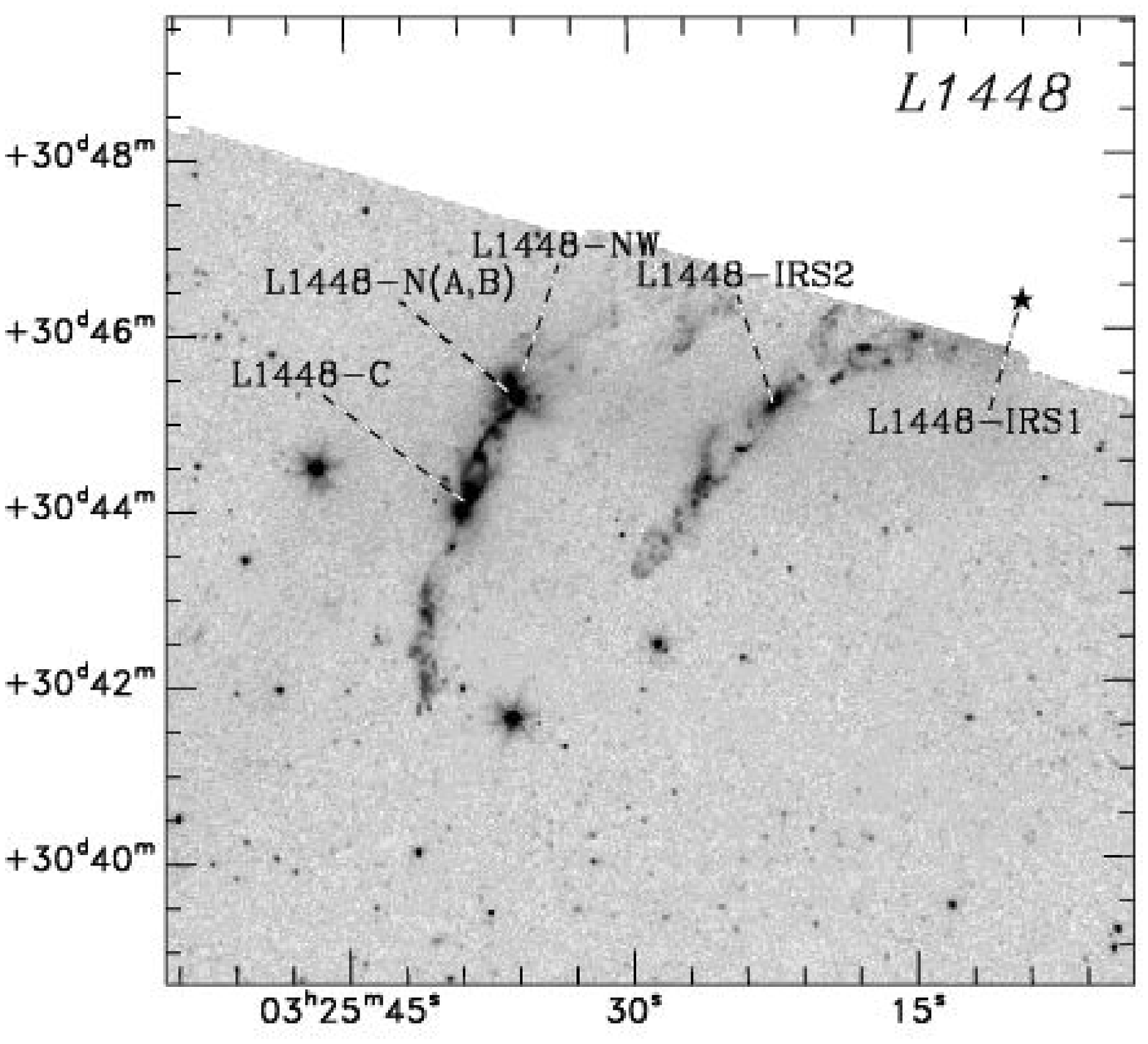}\includegraphics{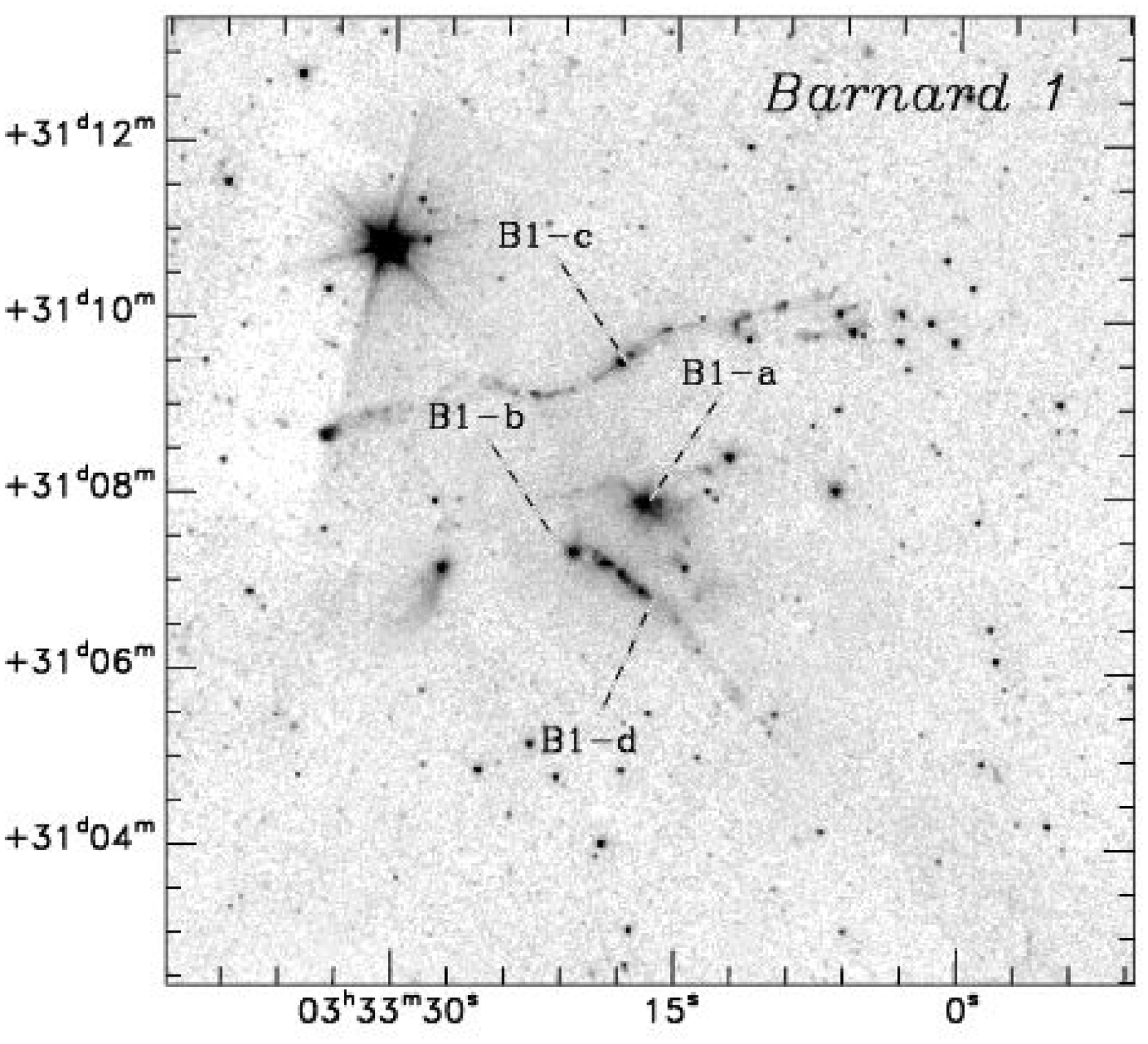}}
\resizebox{\hsize}{!}{\includegraphics{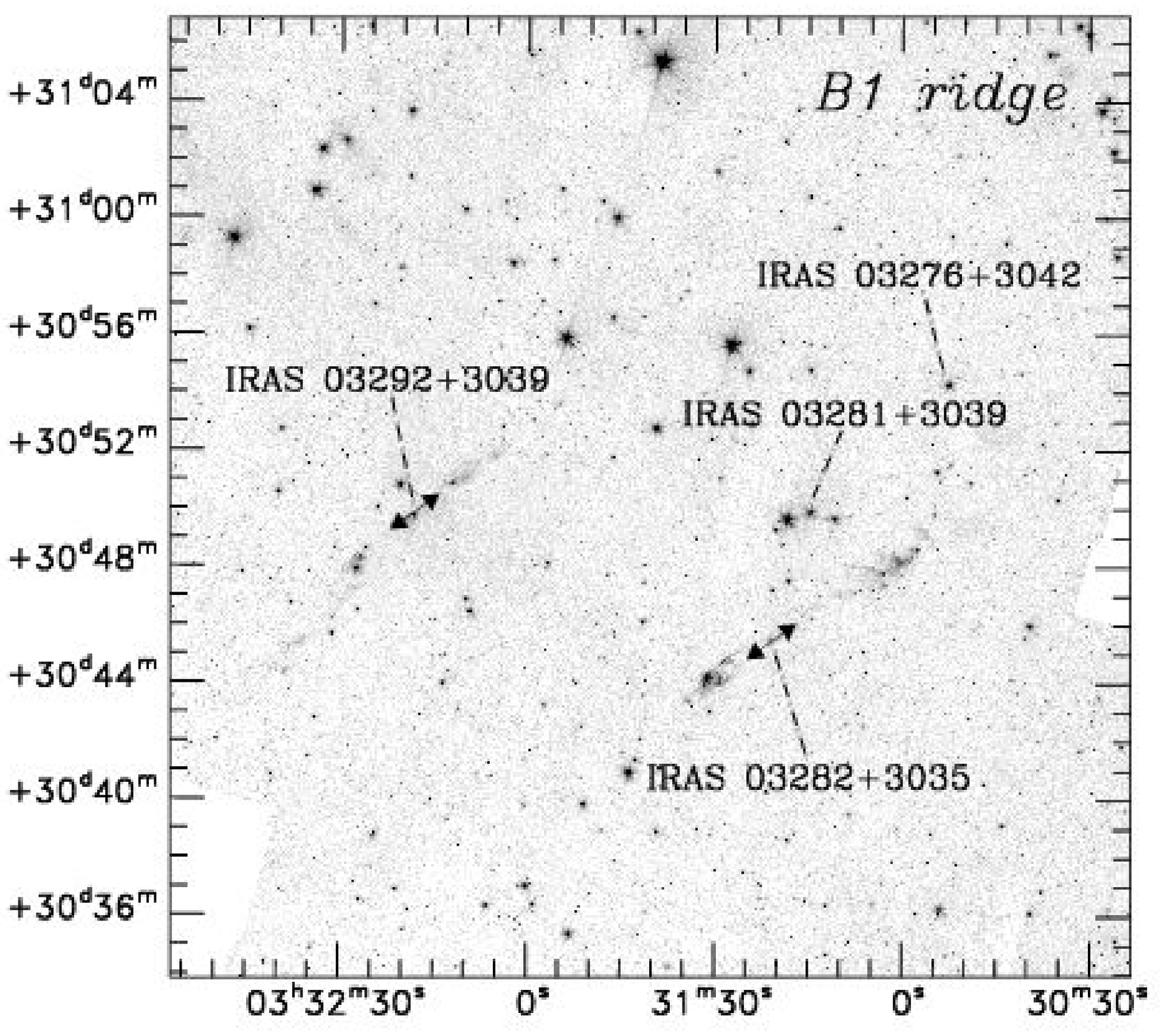}\includegraphics{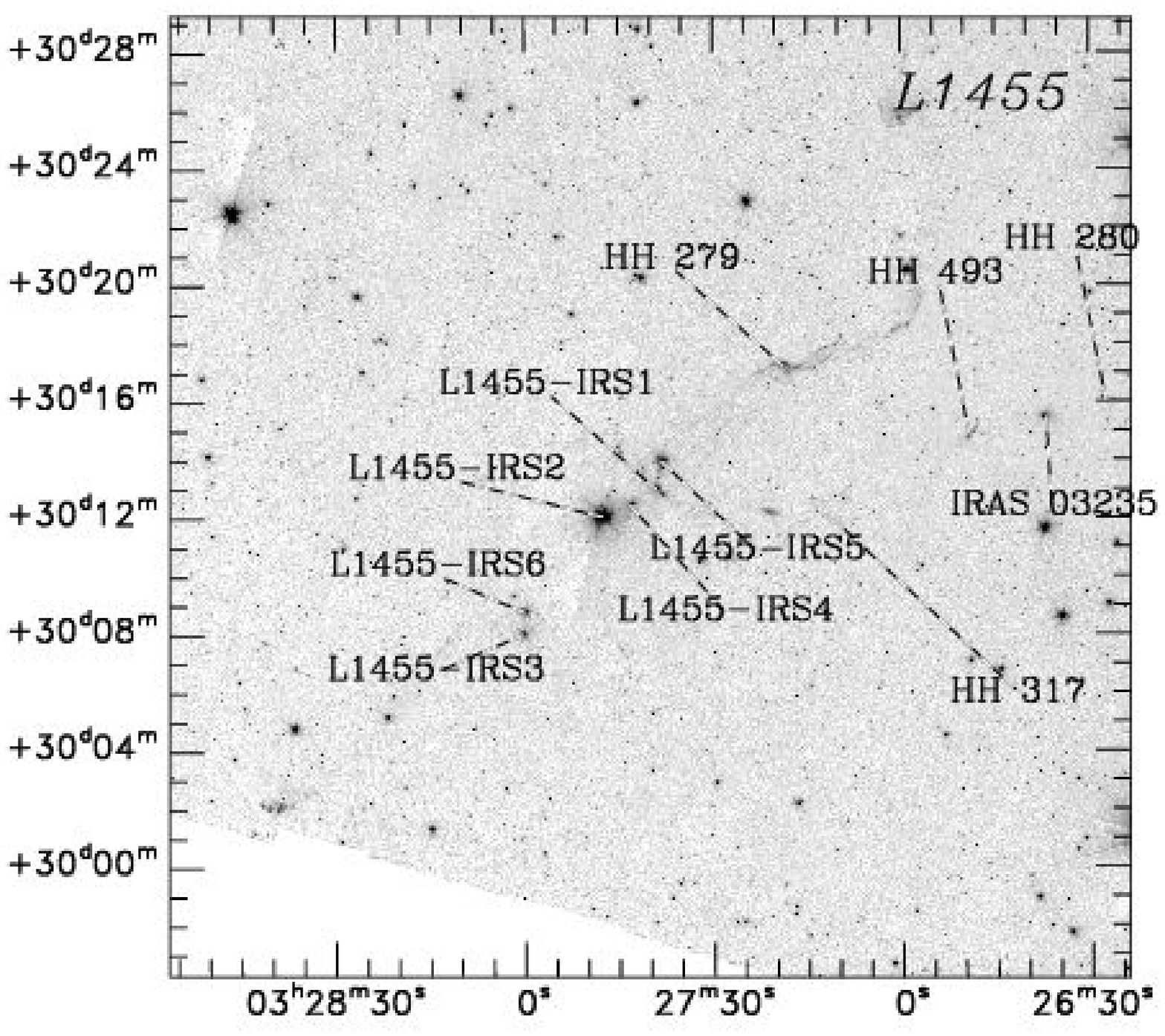}}
\caption{Images of outflows in Perseus in band 2. Upper panels, L1448
  (left) and B1 (right). Lower panels: B1-ridge (left) and L1455
  (right). In the L1448 panel the location of L1448-IRS1 (seen in the
  band 1 and 3 images) is indicated with a black star. In the B1-ridge
  panel the directions of the outflows from IRAS~03292-3039 and
  IRAS~03282-3035 are indicated by black arrows. Similar arrows are
  shown for L1455-IRS1 and L1455-IRS4 in Fig.~\ref{l1455_sharcii}
  where the location of HH~318 also is indicated.}\label{outflowfig}
\end{figure}

Perhaps the most spectacular features in the maps are the prominent
outflows clearly seen in bands 2 and 3 where molecular hydrogen
emission from shocked gas dominates. Fig.~\ref{outflowfig} shows band
2 images of a selection of the outflows outside the two main clusters
in Perseus. Since the outflow activity is thought to be related to the
accretion onto the central star/disk system, it has been suggested
that sources with such strong outflow activity represented the
earliest, most deeply embedded stages of low-mass protostellar
evolution \citep{bontemps96}, the Class 0 stage \citep{andre93}. On
the other hand, the exact same outflows pose a problem for the source
catalogs above. The deeply embedded protostars driving these outflows
likely show extended emission at 4.5 and 5.8~$\mu$m from which it may
be problematic to extract the embedded sources.

\subsection{L1448}
\begin{figure}
\resizebox{\hsize}{!}{\includegraphics{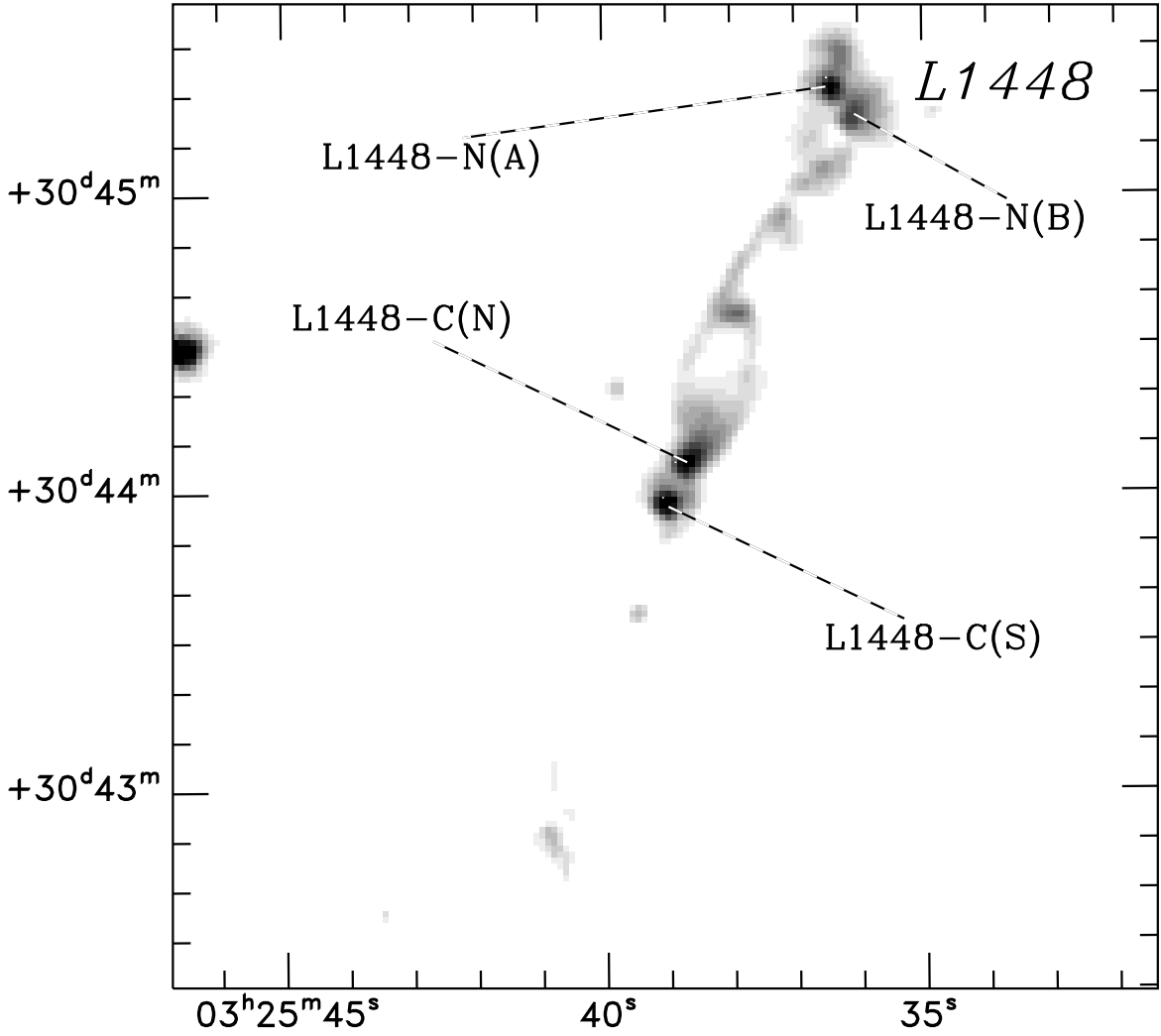}}
\caption{Zoom-in on the IRAC2 image around L1448-C and L1448-N (note
  that the scaling of the image is slightly different from
  Fig.~\ref{outflowfig}). The separation of L1448-N into L1448-N(A)
  and L1448-N(B) and L1448-C into L1448-C(S) and L1448-C(N) is clearly
  seen.}\label{l1448_blowup}
\end{figure}

Fig.~\ref{outflowfig}a shows the outflows in the L1448 region with
known embedded protostars overplotted. The L1448-C source is one of
the best studied Class 0 protostars at a range of wavelengths and is
known to drive a parsec scale outflow in the North-South
direction. This outflow is clearly identified in our maps extending to
the north where the outflow emission makes a break at the positions of
another known Class 0 protostar, L1448-N (or L1448-IRS3). L1448-N is a
binary with a separation of about
7\arcsec\ \citep[e.g.,][]{looney00}. Two sources are seen at the
position of L1448-N(A) and L1448-N(B) with the Eastern, L1448-N(A),
corresponding to the brighter source in the IRAC maps with the
exception of band~1 (Fig.~\ref{l1448_blowup}). Interestingly the
western source, L1448-N(B), is the stronger of the two in 2.7~mm
observations by \cite{looney00}. The Spitzer maps also show some
indication of two fan-shaped outflows emerging from the central source
in the north direction, and separating the point source fluxes is
subject to significant uncertainties. Only faint extended emission is
seen at the location of a third source, L1448-NW (L1448-IRS3C),
northwest of the L1448-N(A) and -N(B). Almost parallel to the L1448-C
outflow, about 8\arcmin\ to the west, the outflow from the Class 0
source L1448-IRS2 \citep[][]{olinger99} is seen. Another outflow is
observed close to L1448-IRS2 in the north-west direction. This outflow
is associated with the L1448-IRS1 source, which is located outside our
mosaic in IRAC bands 2 and 4 but clearly identified in bands 1 and
3. The outflows from L1448 are likely extending even further south:
between L1455 and L1448 two prominent knots associated with shocked
gas are identified at (03:25:51; +30:35:13) and (03:26:04; +30:39:02),
the latter is associated with HH~277, and further south at the
position of HH~278 (03:26:59; +30:25:58) pointing back toward the
L1448 region. Shocks are seen on similar scales pointing back toward
the NGC~1333 region in our maps around the position of HH~746 at
(03:28:32; +30:52:10).

Toward L1448-C, two sources are clearly seen in the IRAC images
separated by about 8\arcsec\ (Fig.~\ref{l1448_blowup}), which we will
refer to as L1448-C(N) (at the position of the millimeter source
L1448-mm) and L1448-C(S). Of these only one source is identified in
all four IRAC bands in our source extraction. The source recognized
from millimeter interferometric observations as the driving source of
the outflow - and which is detected in the millimeter continuum
interferometric measurements - is only detected in IRAC band 3 and
MIPS 24~$\mu$m by the source extraction software even though it is
clearly identified as a separate point source. The reason is likely
the confusion with the extended emission from the outflows in the
other bands. We extract a flux for this source from aperture
photometry in a 2 pixel radius (2.4$''$) aperture with a sky annulus
extending from 2 to 6 pixels (2.4$''$ to 7.2$''$) using the
\emph{aper} procedure from the IDL Astronomy User's
library. L1448-IRS2 also does not make it to the high quality catalog;
it shows extended emission in IRAC2 and IRAC3 and is therefore not
included in our high quality catalog used in the color-color and
color-magnitude diagrams above. The source extractor does identify
this source and it is included in the overall catalog of sources from
c2d.

\subsection{L1455}
The outflows in L1455 pose an interesting puzzle. Previously three
young stellar objects had been identified in this region
\citep{juan93}: L1455-IRS1 (IRAS~03245+3002), L1455-IRS2
(IRAS~03247+3001) and L1455-IRS3 (IRAS~03249+2957). L1455-IRS2 is
driving a North-South outflow and associated with the red nebular
object, RNO~15. An outflow is stretching roughly northwest-southeast
in CO observations \citep{goldsmith84} and is likely associated with a
number of Herbig-Haro objects \citep{bally97}. The driving source of
this outflow is not clear: \cite{juan93} suggested that L1455-IRS1 was
the driving source of the entire complex CO outflow observed by
\cite{goldsmith84}, but \cite{bally97} argued that two separate
outflows had to be responsible for the NW/SE outflow and an outflow
extending in the East-West direction (also responsible for
HH~317). Indeed \cite{davis97} showed from CO 3--2 observations that
L1455-IRS1 was associated with an outflow in the NE/SW direction and
likely responsible for HH~318.

\begin{figure}
\resizebox{\hsize}{!}{\includegraphics{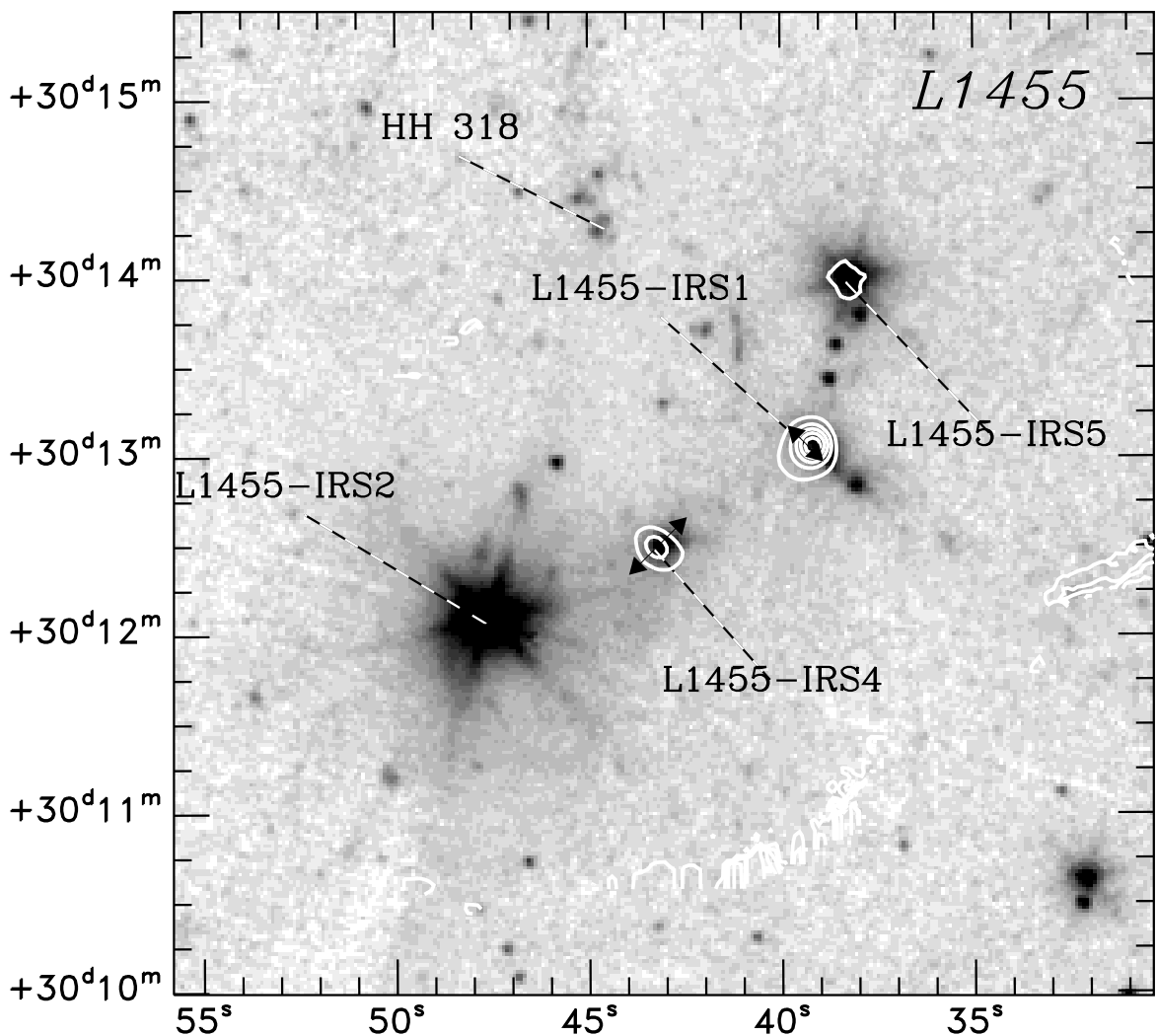}}
\caption{Zoom-in on the IRAC2 image of the central region of L1455
  from Fig.~\ref{outflowfig}. Overlaid in white contours are a
  350~$\mu$m continuum image from the SHARC~II bolometer array at the
  Caltech Submillimeter Observatory (J. Wu, in prep.). The contours
  are plotted at 15\%, 30\%, 45\%, $\ldots$ of the peak flux (at the
  position of L1455-IRS1). The approximate directions of the outflows
  from L1455-IRS1 and L1455-IRS4 are indicated by black
  arrows.}\label{l1455_sharcii}
\end{figure}

In the IRAC2 image the NW/SE outflow is clearly identified together
with a number of the Herbig-Haro objects observed by
\citeauthor{bally97}: HH~279 (03:27:19; +30:17:16), HH~317 (03:27:11;
+30:11:59) and HH~318 (03:27:44.6; +30:14:17). In addition a
nebulosity is seen south of HH~279 delineating the southern part of
the blue-shifted lobe of the CO outflow. About 15\arcmin\ south-west
of the central L1455 cluster another nebulosity is seen, which could
represent the other lobe of the NW/SE outflow. Two additional infrared
sources are observed close to L1455-IRS1 and -IRS2: one in between the
two sources, which we name L1455-IRS4\footnote{Note that
  \cite{froebrich05} erroneously identified a source close to this
  position with the RNO~15 nebula.}  and one slightly northwest of
L1455-IRS1 which we name L1455-IRS5. The close proximity of these
sources suggest an interaction, which may have caused the precession
of the NE/SW outflow as discussed by \cite{davis97}. Of the four
sources, L1455-IRS2 is clearly the strongest at mid-infrared
wavelengths followed by L1455-IRS5. On the other hand L1455-IRS1 and
L1455-IRS4 are associated with the strongest dust continuum peaks in
maps of the (sub)millimeter emission as shown in a map at 350~$\mu$m
taken with the SHARC~II bolometer array at the Caltech Submillimeter
Observatory (Fig.~\ref{l1455_sharcii}; J. Wu et al., (in prep.) - see
also \cite{hatchell05} and \cite{enoch05}). From the color criteria
discussed above, L1455-IRS1 and L1455-IRS4 would both fall in the
Class I group whereas L1455-IRS2 and L1455-IRS5 would be more evolved
Class II objects. It is therefore also likely that L1455-IRS1 and
L1455-IRS4 are responsible for the NE/SW and NW/SE outflows
respectively and their association with the dust continuum peaks
furthermore suggest their deeply embedded natures. L1455-IRS3 is
likewise classified as a Class I source but no continuum peak is seen
toward this source in the SCUBA maps of \cite{hatchell05}
(unfortunately it falls just outside the field observed by
\cite{enoch05}). A companion, L1455-IRS6, is seen 40$''$ north of
L1455-IRS3 with an SED typical of a Class II source. Of the two
candidate embedded protostars, L1455-IRS1 and L1455-IRS4, L1455-IRS1
has full photometry in our high quality catalog from 3.6 to 24~$\mu$m
whereas L1455-IRS4 only is recognized at the two longer wavelengths.

Finally, two knots are seen at (03:26:27; +30:16:01) and (03:26:49;
+30:14:54) in the IRAC2 maps associated with HH~280 and HH~493
distributed symmetrically on a east/southeast-west/northwest axis
through IRAS~03235+3004 which \cite{walawender04} suggested was the
driving source of this outflow. IRAS~03235+3004 is detected as a
candidate Class I object in our catalogs and associated with a strong
continuum peak in the maps of \cite{hatchell05} and \cite{enoch05}.

\subsection{B1 and beyond}
Finally we would like draw the attention to the outflow activity
around the B1 region which has recently received much attention as one
of the very active regions of star formation in the Perseus complex
with a number of submillimeter clumps with embedded protostars and
Herbig-Haro objects \citep{hirano99,matthews02,walawender05b1}. A
number of these prominent outflows are observed in the IRAC data. The
most striking is the precessing outflow clearly associated with the
B1-c source. This source stands out very clearly in the RGB image
(Fig.~\ref{color_rgb}) with a very red color suggesting the embedded
nature of this object. Of the other sources in this region, B1-a shows
extended emission that dominates the source at 3.6~$\mu$m. The source
is therefore not included in the high quality catalog used for the
color-magnitude diagrams above.

The B1-b and B1-d submillimeter cores suggest an interesting
interaction: toward the B1-b position a point source with colors
consistent with an embedded YSO is clearly seen with a jet extending
southwest of the core toward the position of the other submillimeter
core, B1-d. On the other side of this core the jet seems to be making
a break and thereafter extends further south with weaker emission. The
break occurs at the exact position of the B1-d submillimeter core
suggesting some interaction between the outflows and the core. An
outflow extending from a source embedded in the B1-d core can of
course also not be ruled out: no point source is seen in the IRAC
bands but a faint MIPS source is seen toward the position of the
submillimeter core. Previously no outflow had been associated with the
B1-b sources, but this may have been due to the confusion in the
region. Based on maps of the dust continuum emission from SCUBA and
3~mm interferometric measurements from the Nobeyama Millimeter Array,
B1-b was found to be a binary by \cite{hirano99} with a separation of
$\approx$~20$''$, but it does not show up as two separate sources in
the Spitzer images.

In addition to these sources the emission line star, LkH$\alpha$327,
is clearly the strongest source in the IRAC bands located northwest of
the submillimeter cores. LkH$\alpha$327 is identified as a Class II
object by our criteria. Four additional candidate Class I YSOs are
identified southwest of the main concentration of submillimeter
cores. These sources are weak compared to the sources associated with
the central cores, but might be associated with faint dust continuum
emission.

The region southwest of the main B1 core, the B1-ridge, is also
interesting with two strong outflows associated with the two IRAS
sources, IRAS~03282+3035 and IRAS~03292+3039. Both of these sources
are associated with peaks in dust continuum emission in the maps of
\cite{hatchell05} and \cite{enoch05}, but neither is picked up in the
high quality catalog above. A weak IRAC source is associated with
IRAS~03282+3035 which is somewhat extended in a number of the IRAC
bands. IRAS~03292+3039 is not detected in IRAC band 1 by the source
extraction algorithm but recognized as a faint (1.2~mJy) source with
an arcshaped nebulosity in the images. In addition to these two
candidate deeply embedded protostars, two other IRAS sources,
IRAS~03281+3039 and IRAS~03276+3042 are found in this region. The
former was previously classified as a candidate galaxy based on its
IRAS colors \citep{degrijp87}, but has colors consistent with a Class
I YSO in our data. It is located in close proximity to a group of HH
filaments, HH~770, 771 and 772 \citep{walawender05b1} and a group of
sources classified as Class II objects in our analysis. It shows
strong emission at all IRAC and MIPS wavelengths and is detected in
2MASS, but is not associated with any dust continuum peaks. The
latter, IRAS~03276+3042, is likely associated with a candidate Class
II source from our catalog also detected at all wavelengths.

\subsection{Colors of embedded protostars}
\begin{figure}
\resizebox{\hsize}{!}{\includegraphics{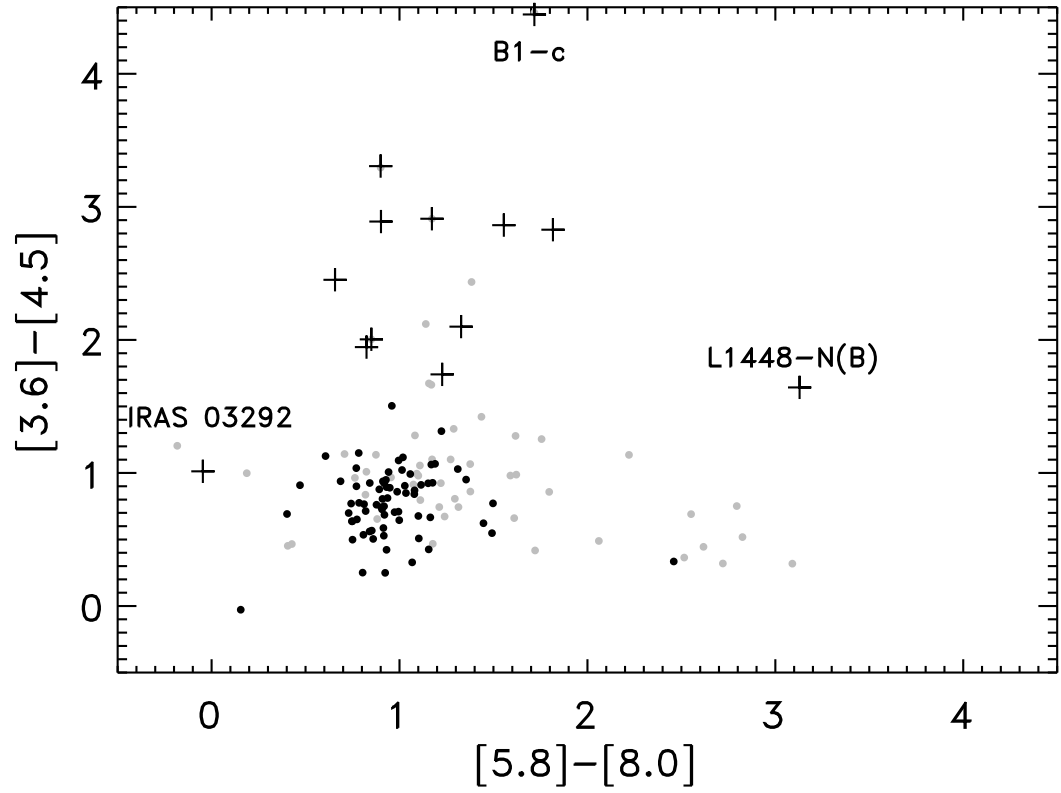}}
\caption{As the left panel of Fig.~\ref{lada_cc} (Class I objects with
  grey and ``flat spectrum'' sources with black symbols) but including the
  candidate deeply embedded protostars (plus signs).}\label{class0cc}
\end{figure}

\begin{figure}
\resizebox{\hsize}{!}{\includegraphics{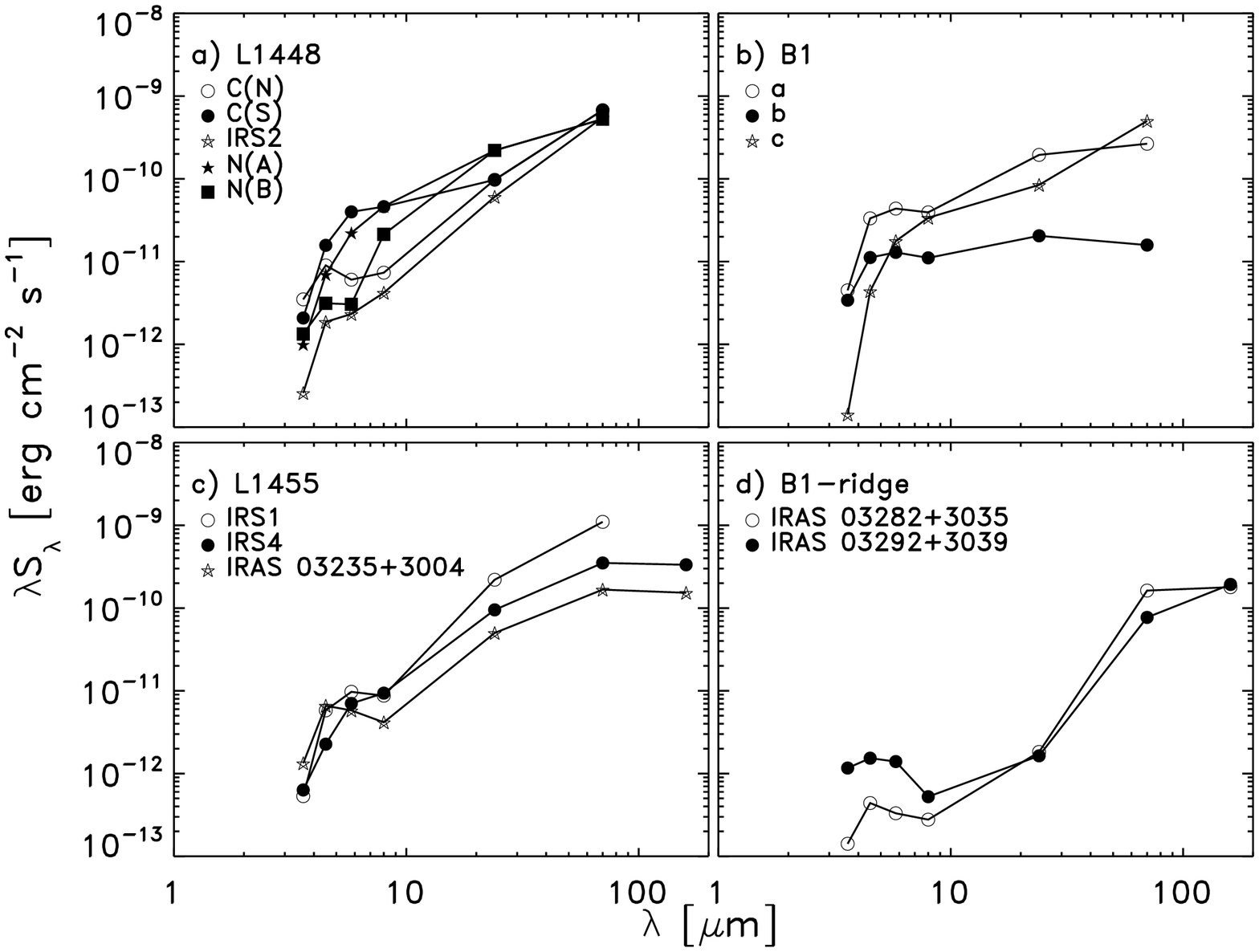}}
\caption{Overview of the 3.6--24$\mu$m SEDs for each of the Class 0
  objects from Table~\ref{embedlist} with additional 70 and 160~$\mu$m
  fluxes from Spitzer/MIPS observations from c2d
  \citep{rebull06}. Note: we assume that L1448-C(N) and L1448-C(S),
  L1448-N(A) and L1448-N(B), respectively each contribute half of
  their observed MIPS fluxes and therefore the plotted value for each
  corresponds to half the number listed in
  Table~\ref{embedlist}.}\label{class0sed}
\end{figure}

As discussed above, each of the known Class 0 objects discussed above
is driving outflows. Table~\ref{embedlist} lists the mid-infrared
measurements of each of the Class 0 objects, together with the
candidate outflow driving sources discussed above. This compilation
does not include sources in the areas covered by the guaranteed time
observations of NGC~1333 and IC~348 discussed by \cite{gutermuth06}
and \cite{lada05}. The detection of mid-infrared counterparts for
these deeply embedded protostars nicely demonstrates the high
sensitivity of the Spitzer observations, since previously only a few
deeply embedded protostars had been measured in the
mid-infrared. Fig.~\ref{class0cc} singles these objects out in the
$[3.6]-[4.5]$ vs. $[5.8]-[8.0]$ diagrams and Fig~\ref{class0sed} shows
the SEDs of each of them. Each of these objects would easily be picked
out as Class I objects from this diagram with IRAS~03292+3039 at the
bluest end of the distribution.  These objects actually are somewhat
separate from the overall population of Class I objects with in
general the reddest colors. It is clear from the diagram that the
Class 0 objects have much redder $[3.6]-[4.5]$ colors than
$[5.8]-[8.0]$ colors. In the models of \cite{allen04} the explanation
is that the 10~$\mu$m silicate absorption feature overlaps with the
IRAC4 (8.0~$\mu$m) band. Objects with progressively more massive
envelopes will have an increasing silicate absorption feature and
therefore less red $[5.8]-[8.0]$ colors, as the 8.0~$\mu$m flux is
absorbed. The objects in Fig.~\ref{class0cc} fall outside the range
spanned by the models of embedded YSOs by \cite{allen04} with even
redder $[3.6]-[4.5]$ colors. Judging from the diagrams of
\cite{allen04} this could imply larger values of the centrifugal
radius (300~AU) which would provide less line of sight absorption,
similar to the modeling results of the Class 0 protostar
IRAS~16293-2422 \citep{iras16293letter}. However, we emphasize that
this is based on simple 1D models not taking the more detailed source
structures into account. \cite{whitney03a} for example showed that the
presence of outflow cavities likewise would lead to bluer colors -
although this effect should be important for both [5.8]-[8.0] and
[3.6]-[4.5] colors according to the models of
\cite{whitney03a}. Additional factors important for these colors could
also be that the (proto)stellar photosphere starts dominating at
shorter wavelengths: this could cause a double peaked structure of the
SED and can explain the SEDs that have a turnover between the IRAC2
and IRAC3 measurements (Fig.~\ref{class0sed}). Also the contribution
from the shocked H$_2$ emission to the 4.5~$\mu$m band or deep
absorption due to ice features included in the IRAC bands will affect
the interpretation of these colors. Combinations of all these effects
could be the explanation for the scatter seen in the $[5.8]-[8.0]$
colors of the Class 0 objects in Fig.~\ref{class0cc}. A more detailed
dust radiative transfer treatment of the full SED from mid-infrared
through (sub)millimeter wavelengths for individual objects is needed
to shed further light on these issues.

The above discussion also illustrates that some of these potentially
very interesting objects may not be singled out directly by the source
extraction procedures due to confusion with nearby sources and
extended H$_2$ emission. Of the 13 known protostars listed in
Table~\ref{embedlist} only 5 (L1448-C(S), L1455-IRS1, B1-b, B1-c and
IRAS~03235+3004) are included in the high quality catalog
(Sect.~\ref{highqualitydef}). It should also be emphasized that the
above discussion is not a complete or an exhaustive discussion of all
the information available about the outflows in the IRAC maps. For
example, information about the shock structure could be obtained by
comparing the ratios between the fluxes in different bands. This is
outside the scope of this paper, however.

\clearpage

\begin{deluxetable}{llllllll}
\rotate
\tabletypesize{\scriptsize}
\tablecaption{Deeply embedded objects/outflow driving sources in Perseus with IRAC and MIPS 24 fluxes. For details about the MIPS data see \cite{rebull06}.}\label{embedlist}
\tablehead{\colhead{Source} & \colhead{RA} & \colhead{DEC} & \colhead{$S$(3.6~$\mu$m)} & \colhead{$S$(4.5~$\mu$m)} & \colhead{$S$(5.8~$\mu$m)} & \colhead{$S$(8.0~$\mu$m)} & \colhead{$S$(24~$\mu$m)} \\
           \colhead{      } & \colhead{  } & \colhead{   } & \colhead{[mJy]}           & \colhead{[mJy]}           & \colhead{[mJy]}           & \colhead{[mJy]}           & \colhead{[mJy]}}
\startdata
L1448-C(N)\tablenotemark{a}       & 03:25:38.9 & +30:44:06.0 & 4.2$\pm$0.4    & 13.5$\pm$0.8  & 11.7$\pm$0.8  & 19.6$\pm$1.2   & 1560$\pm$43  \\
L1448-C(S)                        & 03:25:39.1 & +30:43:58.9 & 2.5$\pm$0.08   & 23.6$\pm$0.6  & 77.2$\pm$0.8  & 123$\pm$1      & \tablenotemark{b}       \\
L1448-IRS2                        & 03:25:22.4 & +30:45:13.6 & 0.31$\pm$0.02  & 2.8$\pm$0.2   & 4.5$\pm$0.2   & 10.2$\pm$0.1   & 485$\pm$4    \\
L1448-N(A)                        & 03:25:36.5 & +30:45:23.2 & 1.2$\pm$0.2    & 10.5$\pm$0.3  & 43.0$\pm$0.5  & 124$\pm$1      & 3530$\pm$104 \\
L1448-N(B)                        & 03:25:36.2 & +30:45:17.1 & 1.6$\pm$0.3    & 4.7$\pm$0.5   & 5.9$\pm$0.6   & 57$\pm$2       & \tablenotemark{b}       \\
B1-a\tablenotemark{c}             & 03:33:16.7 & +31:07:55.1 & 5.4$\pm$0.3    & 50.0$\pm$0.6  & 84.6$\pm$0.5  & 105$\pm$0.9    & 1560$\pm$30  \\
B1-b                              & 03:33:20.3 & +31:07:21.4 & 4.1$\pm$0.04   & 16.8$\pm$0.1  & 25.0$\pm$0.2  & 29.6$\pm$0.2   & 164$\pm$2    \\
B1-c                              & 03:33:17.9 & +31:09:31.8 & 0.17$\pm$0.004 & 6.6$\pm$0.08  & 34.3$\pm$0.2  & 90.3$\pm$0.8   & 676$\pm$10   \\
L1455-IRS1                        & 03:27:39.1 & +30:13:02.8 & 0.64$\pm$0.02  &  8.7$\pm$0.2  & 18.8$\pm$0.3  & 23.3$\pm$0.1   & 1760$\pm$34  \\
L1455-IRS4                        & 03:27:43.3 & +30:12:28.9 & 0.76$\pm$0.03  &  3.4$\pm$0.1  & 13.6$\pm$0.2  & 25.0$\pm$0.2   & 761$\pm$9    \\
IRAS~03235+3004\tablenotemark{c}  & 03:26:37.5 & +30:15:28.2 & 1.6$\pm$0.07   & 9.9$\pm$0.1   & 11.2$\pm$0.08 & 11.1$\pm$0.04  & 400$\pm$3    \\
IRAS~03282+3035                   & 03:31:21.0 & +30:45:30.0 & 0.17$\pm$0.006 & 0.66$\pm$0.02 & 0.64$\pm$0.03 & 0.74$\pm$0.02  & 14.6$\pm$0.2 \\
IRAS~03292+3039                   & 03:32:18.2 & +30:49:46.9 & 1.4$\pm$0.2    & 2.3$\pm$0.1   & 2.7$\pm$0.07  & 1.4$\pm$0.04   & 13.1$\pm$0.1 \\ 
\enddata

\tablenotetext{a}{{L1448-C(N)} is also referred to as L1448-C and
  L1448-mm in the literature, being associated with a strong
  millimeter source in aperture synthesis observations (see discussion
  in text).}
\tablenotetext{b}{{L1448-C(N)} and L1448-C(S) are not separated by the
  source extraction software at MIPS~24~$\mu$m. Likewise for
  L1448-N(A) and L1448-N(B).}

\tablenotetext{c}{B1-a and IRAS~03235+3004 are detected by 2MASS in
  the $K_s$ band: B1-a with a $K_s$ flux of 1.4$\pm$0.1~mJy (upper
  limits of 0.089~mJy in $J$ and 0.38~mJy in $H$) and IRAS~03235+3004
  with a $K_s$ flux of 1.2$\pm$0.08~mJy (upper limits of 0.18~mJy in
  $J$ and 0.50~mJy in $H$). None of the remaining sources have 2MASS
  associations.}
\end{deluxetable}

\clearpage

\section{Summary}
We have presented a survey of 3.86~square~degrees of the Perseus cloud
at 3.6, 4.5, 5.8 and 8.0~$\mu$m using the Spitzer Space Telescope
Infrared Array Camera (IRAC). More than 120,000 sources are identified
across the field. Based on their mid-infrared colors and comparison to
off-cloud and extragalactic control fields, 400 candidate young
stellar objects are identified in this sample. About two thirds of
these YSO candidates are located in two clusters, NGC~1333 and IC~348,
constituting 14\% of the surveyed area. The young stellar objects are
divided into Class I and II objects using the slope of their SEDs from
their 2MASS $K_s$ through IRAC and MIPS 24~$\mu$m fluxes: a clear
difference is seen with few embedded Class I and ``flat spectrum''
YSOs in IC~348 (14\% of the YSO population) compared to NGC~1333 (36\%
of the YSOs) and the remaining, extended cloud (47\% of the
YSOs). This suggests an evolutionary difference, with NGC~1333
consisting of a younger population of YSOs compared to IC~348
(consistent with previous near-infrared observations) but also that a
significant fraction of the current star formation might be going on
outside these clusters in the extended cloud where 61\% of the Class I
objects toward Perseus are found. Finally we discuss a number of the
outflows showing up predominantly through shocked H$_2$ emission in
the IRAC2 maps and identify the mid-infrared counterparts for the
deeply embedded Class 0 objects in the region. All known Class 0
objects in Perseus have mid-infrared counterparts. The Class 0 objects
are found to have very red $[3.6]-[4.5]$ colors compared to
$[5.8]-[8.0]$ colors, but are spread out over a wide range of colors
in this diagram.

\acknowledgments We are grateful to the staff at the Lorentz Center at
Leiden University for hospitality during a three week meeting in July
2005 where a large part of this work was pursued. We thank Mike Dunham
for supplying the SHARC~II image of the L1455 region.  This research
has made use of the SIMBAD database, operated at CDS, Strasbourg,
France. The research of JKJ was supported by NASA Origins Grant
NAG5-13050. Support for this work, part of the Spitzer Legacy Science
Program, was also provided by NASA through contract 1224608, 1230782,
and 1230779 issued by the Jet Propulsion Laboratory, California
Institute of Technology, under NASA contract 1407. Astrochemistry in
Leiden is supported by a NWO Spinoza grant and a NOVA grant. KEY was
supported by NASA under Grant NGT5-50401 ussed through the Office of
Space Science. The SHARC~II data were obtained with support from NASA
Origins grant NNG04GG24G to NJE.

\end{document}